\documentclass[aps,pre,twocolumn,superscriptaddress,10pt]{revtex4-2}
\usepackage[utf8]{inputenc}
\usepackage{color}
\usepackage{amsmath}
\usepackage{amssymb}
\usepackage{xcolor} 
\usepackage{graphicx}
\usepackage{esint}
\usepackage{hyperref}
\usepackage{appendix}
\usepackage{comment}
\makeatletter
\@ifundefined{textcolor}{}
{%
\definecolor{red}{rgb}{0.75, 0.1, 0.1}
 \definecolor{BLACK}{gray}{0}
 \definecolor{WHITE}{gray}{1}
 \definecolor{GREEN}{rgb}{0,1,0}
 \definecolor{BLUE}{rgb}{0,0,1}
 \definecolor{CYAN}{cmyk}{1,0,0,0}
 \definecolor{MAGENTA}{cmyk}{0,1,0,0}
 \definecolor{YELLOW}{cmyk}{0,0,1,0}
}

\makeatother

\begin{document}
\title{Universal trade-off between irreversibility and intrinsic
timescale in thermal relaxation with applications to thermodynamic inference}
\author{Ruicheng Bao}
\email{Contact author: ruicheng@g.ecc.u-tokyo.ac.jp}
\affiliation{Department of Chemical Physics \& Hefei National Laboratory, University
of Science and Technology of China, Hefei 230088, China}
\affiliation{Department of Physics, Graduate School of Science,
The University of Tokyo, Hongo, Bunkyo-ku, Tokyo 113-0033, Japan}

\author{Chaoqun Du}
\affiliation{Department of Chemical Physics \& Hefei National Laboratory, University
of Science and Technology of China, Hefei 230088, China}

\author{Zhiyu Cao}
\affiliation{Department of Chemical Physics \& Hefei National Laboratory, University
of Science and Technology of China, Hefei 230088, China}

\author{Zhonghuai Hou}
\email{Contact author: hzhlj@ustc.edu.cn}
\affiliation{Department of Chemical Physics \& Hefei National Laboratory, University
of Science and Technology of China, Hefei 230088, China}

\begin{abstract}
We establish a general lower bound for the entropy production rate (EPR)
based on the Kullback-Leibler divergence and the Logarithmic-Sobolev
constant that characterizes the time-scale of relaxation. This bound
can be considered as an enhanced second law of thermodynamics. When
applied to thermal relaxation, it reveals a universal trade-off relation
between the dissipation rate and the intrinsic relaxation timescale.
From this relation, a thermodynamic upper bound on the relaxation
time between two given states emerges, acting as an inverse speed
limit over the entire time region. We also obtain a quantum version
of this upper bound, which is always tighter than its classical counterpart,
incorporating an additional term due to decoherence. Remarkably, we
further demonstrate that the trade-off relation remains valid for
any generally non-Markovian coarse-grained relaxation dynamics, highlighting
its significant applications in thermodynamic inference. This trade-off
relation is a new tool in inferring EPRs in molecular dynamics simulations and practical experiments.

\end{abstract}
\maketitle

\section{Introduction}The past twenty years have seen extraordinary
progress in nonequilibrium statistical physics of small systems with
nonnegligible fluctuations. Significant advances include the celebrated
fluctuation theorems \cite{97PRL_Jarzynski,93PRL_FT,94PRE,99PRE_FT,04PRL_EFT,09RMP_FT,10PRL_3FT,12PRE_GFT,12PRL_GFT,05Naturecollin,02PRL_FT,14PRL_FT}
containing all information of the stochastic entropy production (EP), speed
limit in quantum and classical systems \cite{18PRL_SL,22JPCL_SL,20PRL_DissipationTimeUR,22PRX_SL,PRXQuantum_SL,23PRL_TSL,22PRL_TSL,23PRX_Saito},
some refined versions of the second law of thermodynamics \cite{aurell2012refined,15PRL_Roldan,17PRL_Generic,17PRX_Infima,19PRL_Saito,20PRL_Stoppingtime,21PRL_Hasegawa,21PRL_GeometricBound}
and the recently proposed thermodynamic uncertainty relations \cite{15PRL_TUR,16PRL_TUR,17PRE_finiteTUR,20PRR_CZY,21PRR_CTR,21PRX_ImprovedTUR,22PRR_shortTUR,23PRE_BaoTUR,20PRL_Short,17PRL_Todd,doi:10.1063/5.0094211}.
Thermodynamic irreversibility, typically quantified by EP, is key to most of the important theorems and relations mentioned
above. As a central concept in modern thermodynamics, it plays a pivotal role across various fields, including the optimization of heat engines, the design and operation of nano-machines, and the understanding of biological system functions. The thermodynamic irreversibility in transient
processes, which are common in nature and inherently out-of-equilibrium,
has not been studied as extensively as that in stationary processes
\cite{20PRL_FUphill}. 

\normalcolor{Our main focus here is on a crucial and nontrivial
class of transient processes known as thermal relaxation. This fundamental
class of physical processes is ubiquitous in the real world and has
numerous applications across various fields} \cite{23NJP_DPT}. Interestingly,
thermal relaxation phenomena are \normalcolor{complex} and varied
even under Markov approximations. Typical examples are dynamical phase
transitions \cite{22PRL_DynamicalPT,23NJP_DPT}, anomalous relaxation
like Mpemba effect \cite{lu2017nonequilibrium} and asymmetric relaxation
from different directions \cite{20PRL_FUphill,21PRR_Equidistant,21PRR_Asymmetry}.
One of the central quantities in thermal relaxation is its time-scale
of convergence, which has been intensively studied. A well-developed
theory on that is the spectral gap theory, which says that the relaxation
timescale is typically characterized by the spectral gap of the generator
of dynamics in the large time regime. Around the spectral gap, some
important frameworks on metastability \cite{01PRE_Meta,16PRE_Metastable,16PRL_Meta,21PRR_MetaMany,21PRR_Metastable}
and Mpemba effect \cite{lu2017nonequilibrium,klich2019mpemba,kumar2020exponentially,santos2020mpemba,gal2020precooling,baity2019mpemba,21PRL_Qmpemba,Busiello_2021,21PRR_Equidistant,23PRR_bao,bao25prl_accelerate}
have been established, which mainly focus on the large time limit.
However, there are much fewer works concentrating on the \normalcolor{entire}
time region of relaxation processes \cite{2022Symmetrized}. \normalcolor{In
particular, general principles that constrain the behaviors of instantaneous
irreversibility (entropy production rate, EPR) and the relaxation timescale,
applicable at any time during relaxation processes, remain to be investigated.}

In this study, we propose a general lower bound for irreversibility
based on Kullback-Leibler (KL) divergence \cite{cover1999elements}
and Logarithmic-Sobolev (LS) constant \cite{gine2006lectures}, which
is strengthened compared to the standard second law of thermodynamics.
The general bound is then applied to thermal relaxation, revealing
a trade-off relation between the intrinsic timescale and EPR that is valid throughout the entire relaxation process, not just
in the large-time region. A thermodynamic upper bound on the transformation
time between any pair of given states during thermal relaxation follows
from the trade-off, which we \normalcolor{term} the inverse speed
limit. \normalcolor{More importantly, we theoretically and numerically
show that our trade-off relation holds even for generally non-Markovian
coarse-grained dynamics, significantly broadening the applicability
of the relation. It can aid in the design} \normalcolor{of real-world}
rapid relaxation processes, which are desirable in numerous situations
\cite{21PRL_Qmpemba}. 

\normalcolor{A key distinction between our findings and previous
results is that we provide an experimentally feasible lower bound for the instantaneous
EPR. This contrasts with
prior related results, which mainly focus on lower bounds for
the EP over a time interval during relaxation
\cite{18PRL_SL,22PRL_TSL,19PRL_Saito}, thus representing a different aspect
of nonequilibrium phenomena \cite{deffner_generalized_2010}. Additionally,
the intrinsic timescale considered here is a property of the underlying
dynamics, characterized by the spectral properties of the dynamical generator and independent of the initial and final distributions of the system. In comparison, timescales incorporated
in previous findings, such as speed limits, depend on those
distributions.}

\normalcolor{The trade-off relation for coarse-grained dynamics is fundamentally new and serves as a valuable tool for thermodynamic inference, a crucial task in nonequilibrium statistical physics \cite{Seifert19inference}. We apply this coarse-grained trade-off to infer the EPR in relaxation processes of complex systems, where only coarse-grained observations are feasible, typical in experimental settings. This lower bound complements previous thermodynamic inference results, which are primarily limited to stationary processes \cite{roldan_estimating_2010,martinez_inferring_2019,skinner_estimating_2021,ro_model-free_2022,MES22PRX,HDPR22PRX,MDS23PRL,blom_milestoning_2024,di_terlizzi_variance_2024}. Although some studies have focused on non-stationary dynamics \cite{Shiraishi15PRE,LKP23multi,KA22PRE,otsubo2022estimating,Ohga_prr}, these methods often require substantial trajectory data, complex procedures, and assume Markovianity, making them impractical for non-Markovian coarse-grained dynamics. In contrast, our methodology is applicable to highly coarse-grained dynamics with very few coarse-grained states, without requiring trajectory data, relying solely on the statistics of coarse-grained states, which are easier to obtain in practice. We demonstrate our method with coarse-grained data from molecular dynamics simulations, where transitions or currents are undetectable and heat dissipation cannot be measured directly. This demonstration highlights the potential applicability of the trade-off relation for inferring EPR in real experiments.}

\section{A general lower bound for EPR}We
are considering a system with $N$ states coupled to a heat bath with
inverse temperature $\beta=1/(k_{B}T)$, though the generalization
of our results to multiple heat baths is straightforward. The dynamics
of the probability of the system being in state $i$ at time $t,$
$p_{i}(t)$, is described by a master equation
\begin{equation}
\frac{\text{d}}{\text{d}t}p_{i}(t)=\sum_{j=1}^{N}\left[k_{ij}(t)p_{j}(t)-k_{ji}(t)p_{i}(t)\right],\label{ME}
\end{equation}
where $k_{ij}(t)$ denotes the transition rate from state $j$ to
state $i$ at time $t$. The master equation can be rewritten in a
more compact matrix form as $\frac{\text{d}}{\text{d}t}\boldsymbol{p}(t)=\mathcal{L}(t)\boldsymbol{p}(t)$,
where $\boldsymbol{p}(t)=[p_{1}(t),p_{2}(t),...,p_{N}(t)]^{\text{T}}$
and $\mathcal{L}_{ij}(t)=k_{ij}(t)-\delta_{ij}\sum_{l}k_{li}(t)$
is the stochastic matrix (strictly speaking, $\mathcal{L}$ is an
operator) at time $t$. The stochastic matrix changes over time due
to external protocols. In this work, we focus on both cases where
the detailed balance condition $k_{ij}(t)\pi_{j}^{t}=k_{ji}(t)\pi_{i}^{t}$
holds for all pairs of $i$, $j$ at any time $t$, and when it does
not, where \normalcolor{$\pi_{i}^{t}$ is the (instantaneous) stationary
distribution at time $t$ for state $i$.} We denote $\boldsymbol{\pi}_{t}=[\pi_{1}^{t},...,\pi_{N}^{t}]^{\text{T}}$,
in which $\boldsymbol{\pi}_{t}$ is defined as the stationary state
will be reached if the stochastic matrix is frozen at time $t$. \normalcolor{When
the detailed balance condition holds, $\boldsymbol{\pi}_{t}$ will
be an (instantaneous) equilibrium state $\boldsymbol{p}_{t}^{\text{eq}}$
whose entries are $p_{t,i}^{\text{eq}}=e^{-\beta E_{i}(t)}/Z$, with
$E_{i}(t)$ being the instantaneous energy of state $i$ at time $t$
and $Z$ the normalization constant.} The KL divergence, which quantifies
the difference between two probability distributions, is defined as
$D[\boldsymbol{p}^{a}\vert\vert\boldsymbol{p}^{b}]\equiv\sum_{i}p_{i}^{a}\ln(p_{i}^{a}/p_{i}^{b})$.
For any continuous-time Markov processes obeying the master equation
(\ref{ME}) with an instantaneous \normalcolor{equilibrium} distribution
$\boldsymbol{p}_{t}^{\text{eq}}$ at time $t$, we demonstrate that
\cite{SupplementalMaterial}
\begin{equation}
\frac{\text{d}}{\text{d}\tau}D[\boldsymbol{p}(\tau)\vert\vert\boldsymbol{p}_{t}^{\text{eq}}]|_{\text{\ensuremath{\tau=t}}}\leq-4\lambda_{\text{LS}}(t)D[\boldsymbol{p}(t)\vert\vert\boldsymbol{p}_{t}^{\text{eq}}],\label{KLI}
\end{equation}
where $\lambda_{\text{LS}}(t)$ is a positive real number determined
by $\mathcal{L}(t)$. Further, without detailed balance condition,
we still have a similar inequality $\frac{\text{d}}{\text{d}\tau}D[\boldsymbol{p}(\tau)\vert\vert\boldsymbol{\pi}_{t}]|_{\text{\ensuremath{\tau=t}}}\leq-2\lambda_{\text{LS}}(t)D[\boldsymbol{p}(t)\vert\vert\boldsymbol{\pi}_{t}]$,
where a factor $1/2$ is multiplied on the right hand side. Before
proceeding, we denote $\langle f,g\rangle_{\pi}\equiv\sum_{i}fg_{i}^{\dagger}\pi_{i}$
the inner product induced by the stationary distribution $\boldsymbol{\pi}$
(may be instantaneous). 

The positive real number $\lambda_{\text{LS}}(t)$ in Eq. (\ref{KLI})
is the LS constant \cite{gine2006lectures} corresponding to the stochastic
matrix $\mathcal{L}(t)$, whose definition is 
\begin{equation}
    \lambda_{\text{LS}}\equiv\inf_{\text{Ent}(f)\neq0}\frac{\text{Re}\langle-\mathcal{L}f,f\rangle_{\pi}}{\text{Ent}(f)},
\end{equation}
where $\text{Ent}(f)$ is an entropy-like quantity defined as $\text{Ent}(f)=\sum_{i=1}^{N}\vert f_{i}\vert^{2}\ln\left(\frac{\vert f_{i}\vert^{2}}{\langle f,f\rangle_{\pi}}\right)\pi_{i}$
and $f$ is any function in the state space of the system. 

According to the stochastic thermodynamics, the average EP
rate $\dot{\sigma}(t)$ at time $t$ in this system is ($k_{B}$ is
set to be $1$) \cite{seifert2012stochastic}
\begin{equation}
\dot{\sigma}(t) =\sum_{i,j}k_{ij}(t)p_{j}(t)\ln\frac{k_{ij}(t)p_{j}(t)}{k_{ji}(t)p_{i}(t)}.
\end{equation}
If the stochastic matrix satisfies the detailed balance condition (i.e. the Markov process in focus is reversible),
$\dot{\sigma}(t)$ is related to the KL divergence between the current
distribution and the instantaneous equilibrium distribution \normalcolor{$\boldsymbol{p}_{t}^{\text{eq}}$}
as $\dot{\sigma}(t)=-\partial_{\tau}D[\boldsymbol{p}(\tau)\vert\vert\boldsymbol{p}_{t}^{\text{eq}}]|_{\tau=t}$
\cite{10PRE_Master,21PRL_Bitreset}. Combining this with Eq. (\ref{KLI})
leads to 
\begin{equation}
\dot{\sigma}(t)\geq4\lambda_{\text{LS}}(t)D\left[\boldsymbol{p}(t)\vert\vert\boldsymbol{p}_{t}^{\text{eq}}\right].\label{LBEPR}
\end{equation}
This general lower bound for the EPR at any given
time is our first main result. The bound will always be positive unless
the system is in an equilibrium state, since $\lambda_{\text{LS}}(t)$
is always positive \cite{gine2006lectures}\normalcolor{, }which
makes it generally stronger than the conventional second law.\normalcolor{{}
It also shows that the possible EPR increases
as the system deviates further from the instantaneous equilibrium
state.}

\normalcolor{In the absence of the detailed balance condition,
a similar lower bound for the non-adiabatic EPR
(also named as Hatano-Sasa EP) can be obtained as
}
\begin{equation}
\dot{\sigma}^{\text{na}}(t)\geq2\lambda_{\text{LS}}(t)D\left[\boldsymbol{p}(t)\vert\vert\boldsymbol{\pi}_{t}\right],\label{LBNEPR}
\end{equation}
\normalcolor{where} the definition of $\dot{\sigma}^{\text{na}}(t)$
is given by $\dot{\sigma}^{\text{na}}(t)=-\sum_{i}\dot{p}_{i}(t)\ln\frac{p_{i}(t)}{\pi_{i}^{t}}$
and \normalcolor{the relation $\dot{\sigma}^{\text{na}}(t)=-\partial_{\tau}D[\boldsymbol{p}(\tau)\vert\vert\boldsymbol{\pi}_{\tau}]|_{\tau=t}$
has been used \cite{23PRR_Ito}.} Since the total EP
rate satisfies $\dot{\sigma}(t)\geq\dot{\sigma}^{\text{na}}(t)$,
the bound can still serve as a stronger second law, i.e., $\dot{\sigma}(t)\geq\dot{\sigma}^{\text{na}}(t)\geq2\lambda_{\text{LS}}(t)D\left[\boldsymbol{p}(t)\vert\vert\boldsymbol{\pi}_{t}\right]\geq0$.

In what follows, we focus on an important application of our lower
bounds in thermal relaxation processes, where the stochastic matrix\normalcolor{{}
$\mathcal{L}$ becomes} time-independent and\normalcolor{{} $\lambda_{\text{LS}}$
is a constant uniquely determined by $\mathcal{L}$.} Nonetheless,
in Sec. II of \cite{SupplementalMaterial}, we also\normalcolor{{}
present} another application of Eq. (\ref{LBEPR}) in a system with
time-dependent dynamics, where the stochastic matrix is periodically
switching \cite{busiello2021dissipation,23PRE_Lu}. In that example,
we show that our lower bound can \normalcolor{help in recovering}
part of the ``hidden'' EPR \cite{16PRL_Tang}
of an effective equilibrium state.

\section{Trade-off relation for thermal relaxation}\normalcolor{In
thermal relaxation, $\lambda_{\text{LS}}$ is related to the intrinsic
timescale $\tau^0_{\text{rel}}\equiv\inf_{\tau}\left\{ \sup_{i}\left\Vert \frac{\boldsymbol{p}(\tau|p_{j}(0)=\delta_{ij})}{\boldsymbol{\pi}}-\mathbb{I}\right\Vert _{2}\leq1/e\right\} $,
which characterizes the slowest (dominant) mode of relaxation} \normalcolor{and
is independent of the initial and final distributions.} \normalcolor{$\tau^0_{\text{rel}}$
satisfies $\tau^0_{\text{rel}}\geq1/(2\lambda_{\text{LS}})$. 

Operationally, $\lambda_{\text{LS}}$
can be measured via another expression:
\begin{equation}
    \frac{1}{2\lambda_{\text{LS}}}\equiv\tau^1_{\text{rel}}\overset{t\gg1}{\simeq}\frac{-t}{2\ln D\left[\boldsymbol{p}(t)\vert\vert\boldsymbol{p}^{\text{eq}}\right]}.
\end{equation}
Here, $\tau^1_{\text{rel}}$ is the measurable relaxation timescale. This may
require more statistical data for precise measurement but is more
adaptable to non-Markovian coarse-grained dynamics. For both definitions,
Eq. (\ref{LBEPR}) leads to}

\normalcolor{
\begin{align}
\sigma_{[0,\tau]} & \geq\left(1-e^{-2\tau/\tau^{0,1}_{\text{rel}}}\right)D\left[\boldsymbol{p}(0)\vert\vert\boldsymbol{p}^{\text{eq}}\right],\ \tau\geq0,\label{tradeoff1}
\end{align}
}and \normalcolor{
\begin{equation}
\dot{\sigma}(t)\tau^{0,1}_{\text{rel}}\geq2D\left[\boldsymbol{p}(t)\vert\vert\boldsymbol{p}^{\text{eq}}\right].\label{tradoff2}
\end{equation}
}Here, $\boldsymbol{p}(t)$ is the distribution at time $t$, $\boldsymbol{p}^{\text{eq}}$
is the equilibrium distribution of $\mathcal{L}$ and the entropy
production reads $\sigma_{[0,\tau]}=\int_{0}^{\tau}\dot{\sigma}(t)\text{d}t=D[\boldsymbol{p}(0)\vert\vert\boldsymbol{p}^{\text{eq}}]-D\left[\boldsymbol{p}(\tau)\vert\vert\boldsymbol{p}^{\text{eq}}\right]$.\normalcolor{{}
Eq. (\ref{tradeoff1}) and (\ref{tradoff2}) reveal close connections
between EP (rate) and intrinsic relaxation timescale,}
which is our second main result. These two inequalities hold for any
$t>0$ ($\tau>0$), and they are saturated at the large time limit
$t,\ \tau\rightarrow\infty$. Eq. (\ref{tradeoff1}) also saturate
at the small time limit $\tau\rightarrow0$ when both sides equal
zero. \normalcolor{Eq. (\ref{tradoff2}) rigorously shows that the minimal possible dissipation
rate in thermal relaxation increases as the distance from equilibrium
grows, a fact that was not explicitly known before. Previously, it
was only proven that the accumulated EP from time
0 to $t$ in an irreversible process can be lower bounded by the KL
divergence from the state at $t$ to the equilibrium state \cite{suri_dissipation_2009,deffner_generalized_2010,esposito_second_2011}.
}

\begin{figure}
\begin{centering}
\includegraphics[width=1\columnwidth]{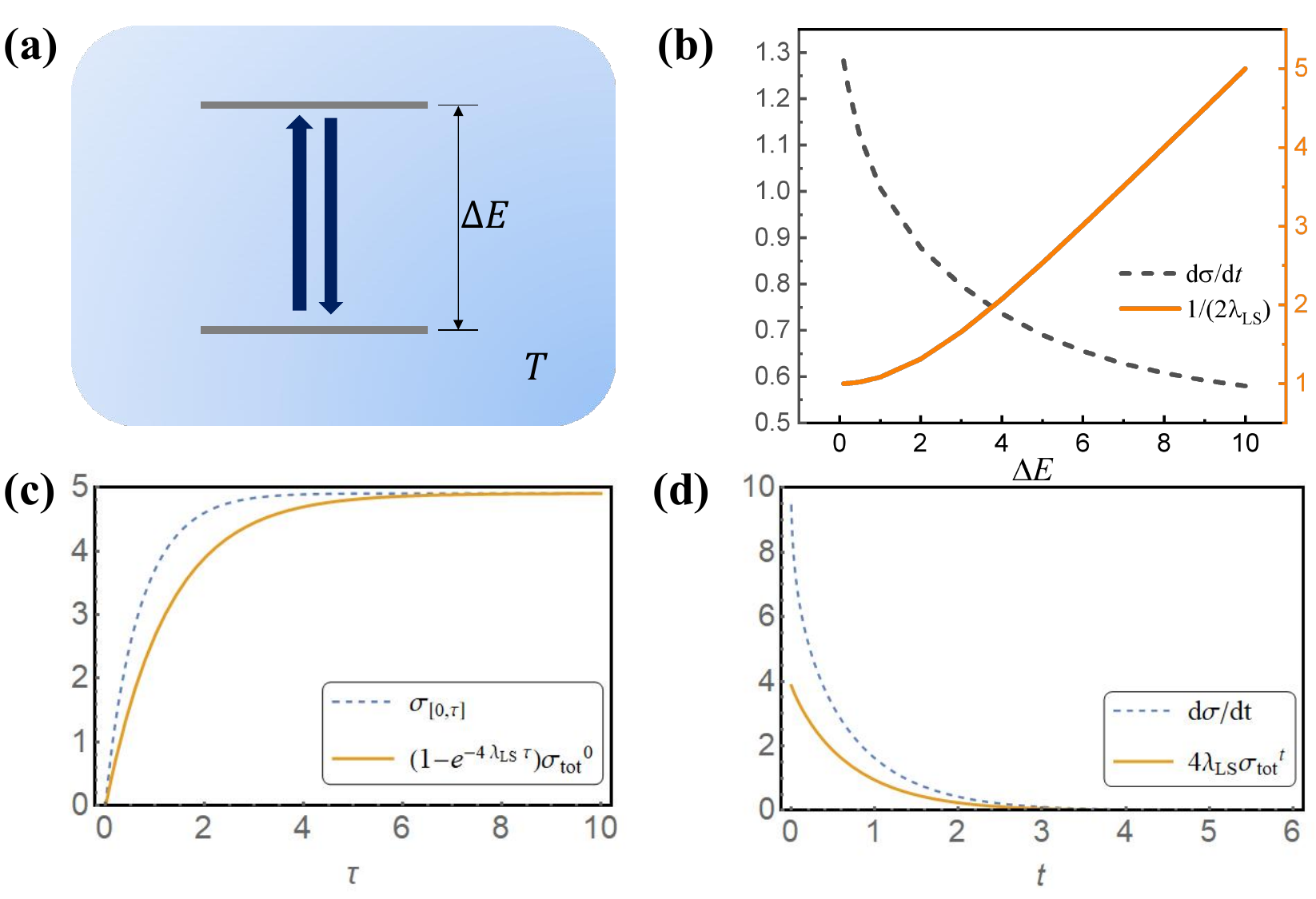}
\par\end{centering}
\caption{Illustration of the trade-off relation and the lower bound for entropy
production in a two-state mode. Here, we set $\beta=1/(k_{B}T)=1$.
(a) The two-state model coupled to a heat reservoir with temperature
$T$. (b) The trade-off relation between the EPR
$\dot{\sigma}(t)$ at $t=0$ and the relaxation time scale $1/(2\lambda_{\text{LS}})$
for relaxation processes with different $\Delta E$, where the distance
to equilibrium $\sigma_{\text{tot}}^{t}\equiv D\left[\boldsymbol{p}(t)\vert\vert\boldsymbol{p}^{\text{eq}}\right]$
is fixed to be $0.5$. (c) and (d) We demonstrate Eq. (\ref{tradeoff1})
and (\ref{tradoff2}) for this model, in which $\Delta E=5.0$ and
the initial distribution is chosen to be $(p_{u},p_{d})=(0.99,0.01)$.}

\label{fig1}
\end{figure}

To illustrate the results, we take a two-state model, which may be
used to model a single spin or a qubit, as an example. As shown in
Fig. \ref{fig1} (a), the model system is comprised of an up state
$u$ with energy $E_{u}$ and an down state $d$ with energy $E_{d}$,
and it is coupled to a heat bath with temperature $T$. The energy
difference between two states is $\Delta E=E_{u}-E_{d}>0$. The transition
rates from $u$ to $d$ and from $d$ to $u$ are given by $k_{u\rightarrow d}=e^{\beta\Delta E}/(1+e^{\beta\Delta E})$
and $k_{d\rightarrow u}=1/(1+e^{\beta\Delta E})$ respectively. Under
this setting, the stationary distribution will be an equilibrium one
$[p_{u},p_{d}]^{\text{T}}=[1/(1+e^{\beta\Delta E}),e^{\beta\Delta E}/(1+e^{\beta\Delta E})]^{\text{T}}$.
The LS constant in this case can be exactly computed as \cite{gine2006lectures}
$\lambda_{\text{LS}}=\frac{\tanh(\frac{\beta\Delta E}{2})}{\beta\Delta E}$.
In Fig. \ref{fig1} (b), a trade-off relation between $\tau_{\text{rel}}=1/(2\lambda_{\text{LS}})$
and $\dot{\sigma}(t)$ is demonstrated when $D\left[\boldsymbol{p}(t)\vert\vert\boldsymbol{p}^{\text{eq}}\right]$
is fixed. Fig. \ref{fig1} (c) and (d) shows that two relations (\ref{tradeoff1})
and (\ref{tradoff2}) are valid for any time.

\normalcolor{Additionally, we obtain another bound related to the
spectral gap $\lambda_{g}$ of $\mathcal{L}$ as 
\begin{equation}
\dot{\sigma}(t)\geq4C\lambda_{g}D\left[\boldsymbol{p}(t)\vert\vert\boldsymbol{p}^{\text{eq}}\right],\label{bound2}
\end{equation}
where $C=(1-2\pi_{\star})/\ln(1-\pi_{\star}/\pi_{\star})$. The relation
between EP and spectral gap is of broad interest \cite{oberreiter_universal_2022,ohga_thermodynamic_2023,kolchinsky_thermodynamic_2024,van_vu_dissipation_2024}.
Eq. (\ref{bound2}) is a complementary relation to previous results
on stationary EP.}

In Sec. VII of \cite{SupplementalMaterial}, we further show generalizations
of the trade-off relation to discrete-time Markov processes and continuous-space
Markov processes. 

\section{Inverse speed limit}A \normalcolor{corollary} of
Eq. (\ref{tradoff2}) is an inverse speed limit:
\begin{align}
\tau & \leq\frac{1}{4\lambda_{\text{LS}}}\ln\left\{ \frac{D\left[\boldsymbol{p}(0)\vert\vert\boldsymbol{p}^{\text{eq}}\right]}{D\left[\boldsymbol{p}(0)\vert\vert\boldsymbol{p}^{\text{eq}}\right]-\sigma_{[0,\tau]}}\right\} ,\label{ISL}
\end{align}
which gives the upper bound for the time $\tau$ of the relaxation
from an initial distribution $\boldsymbol{p}(0)$ to a target distribution
$\boldsymbol{p}(\tau)$. \normalcolor{Here, the EP
$\sigma_{[0,\tau]}$ for the state transformation from $\boldsymbol{p}(0)$
to $\boldsymbol{p}(\tau)$ can be interpreted as the distance
between these two states, as it is monotonic in time during relaxation.
Thus, the maximal time that the system takes to relax through such
a distance $\sigma_{[0,\tau]}$ is given by the inverse speed limit
(\ref{ISL}). The system should initially be farther from equilibrium
so that the transformation time $\tau$ can be shorter. The upper
bound still holds when the detailed balance condition is not satisfied.
The only difference is that the EP should be replaced
with the non-adiabatic EP $\sigma^{\text{na}}$ and
a factor $1/2$ should be multiplied on the right.}

\normalcolor{The above relation (\ref{ISL}) can be generalized
to open quantum systems described by the Lindblad master equation.
The population in open quantum system is defined as $P_{n}(t)\equiv\langle n\vert\rho_{t}\vert n\rangle$,
where $\rho_{t}$ is the density matrix and $|n\rangle$ is the $n$-th
energy eigenstate. Then, one can show that an upper bound on the transformation
time $\tau$ that is tighter than its classical counterpart is given
by
\begin{equation}
\tau\leq\frac{1}{4\lambda_{\text{LS}}}\ln\left\{ \frac{D[\boldsymbol{P}(0)\vert\vert\boldsymbol{P_{\beta}}]}{D[\boldsymbol{P}(0)\vert\vert\boldsymbol{P_{\beta}}]-(\sigma_{[0,\tau]}-\Delta A)}\right\} ,
\end{equation}
with $\boldsymbol{P}(0)$ being the population vector at initial time,
$P_{\beta}$ being the population vector for Gibbs state, $\Delta A\equiv A(0)-A(\tau)\geq0$
and $A(t)$ being the asymmetry defined as $A(t):=D(\rho_{t}||\rho_{t}^{\text{d}})$.
Here, $\rho_{t}^{\text{d}}$ is the fully decohered version of $\rho_{t}$.
This result implies that relaxation may be accelerated by quantum
coherence.} 

\section{\normalcolor{Trade-off relation for arbitrary coarse-grained dynamics
and thermodynamic inference}}

\normalcolor{Measuring the EPR at the coarse-grained
level is challenging due to experimental resolution limitations and
the large amount of data needed for convergence. Remarkably, our trade-off
relation can be generalized to arbitrarily coarse-grained relaxation
dynamics, allowing it to be further applied to estimate EP
when only coarse-grained observations are feasible. The trade-off
relation for coarse-grained dynamics reads 
\begin{equation}
\dot{\sigma}(t)\geq\frac{2}{\tau_{\text{rel}}^{1,\text{CG}}}D\left[\mathcal{P}(t)\vert\vert\mathcal{P}^{\text{eq}}\right],\label{CGLB1}
\end{equation}
where $\mathcal{P}(t)$ and $\mathcal{P}^{\text{eq}}$ are the probability
distributions for coarse-grained states and $\tau_{\text{rel}}^{1,\text{CG}}\overset{t\gg1}{\simeq}\frac{-t}{2\ln D\left[\mathcal{P}(t)\vert\vert\mathcal{P}^{\text{eq}}\right]}$
is the measurable relaxation timescale measured at the coarse-grained level (another
definition $\tau_{\text{rel}}^{0,\rm CG}\equiv\inf_{\tau}\left\{ \sup_{i}\left\Vert \frac{\mathcal{P}(\tau|\mathcal{P}_{j}(0)=\delta_{ij})}{\mathcal{P}^{\text{eq}}}-\mathbb{I}\right\Vert _{2}\leq1/e\right\} $
works when there is timescale separation). The lower bound (\ref{CGLB1})
is our third main result. Physically, we have that $\tau_{\text{rel}}^{\text{CG}}\leq\tau_{\text{rel}}$
because equilibrium of microscopic dynamics implies the convergence
of macroscopic coarse-grained dynamics, but not vice versa. These
inequalities must hold when $\tau_{\text{rel}}^{\text{CG}}\sim\tau_{\text{rel}}$
\cite{SupplementalMaterial}, which is a criterion of good coarse-graining
\cite{yang_slicing_2023}. The coarse-grained version of Eq. (\ref{LBNEPR})
similarly holds. Note that even when there is timescale separation
and $\tau_{\text{rel}}^{\text{CG}}\sim\tau_{\text{rel}}$, the resulting
coarse-grained dynamics can still be non-Markovian \cite{Godec24emergence}.
We also provide theoretical justification for the validity of the
bound in general cases \cite{SupplementalMaterial}, where $\tau_{\text{rel}}^{\text{CG}}$
can be much smaller than $\tau_{\text{rel}}$. We prove a stronger
lower bound without assuming a timescale separation and argue that
it gives the desired experimentally feasible bound. To our knowledge,
there are no similar results like (\ref{CGLB1}) that can infer the
EPR in arbitrarily coarse-grained relaxation dynamics
without knowing the model details. }

\begin{figure}
\begin{centering}
\includegraphics[width=1\columnwidth]{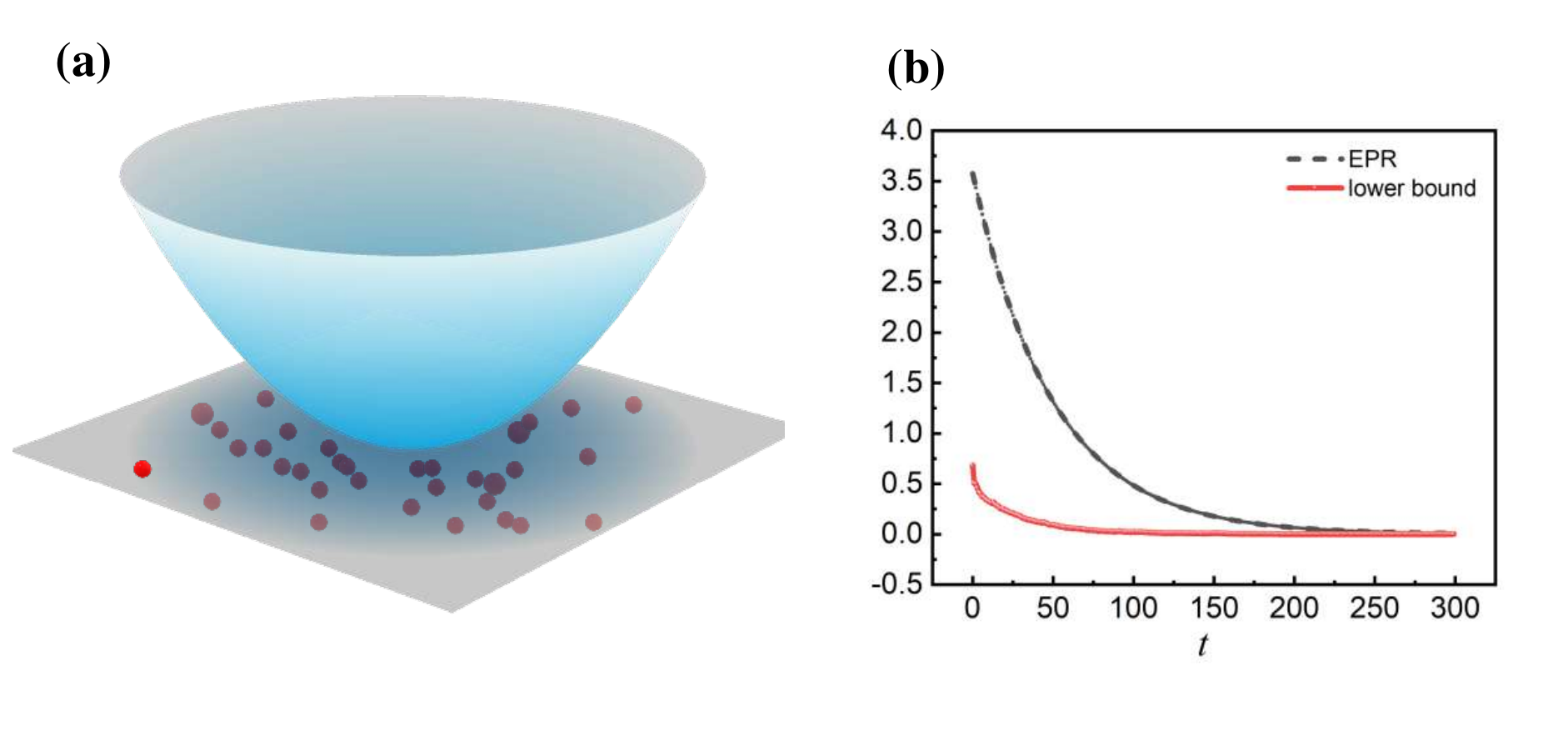}
\par\end{centering}
\caption{\normalcolor{Application of the coarse-grained trade-off relation
to molecular dynamics simulation. (a) The two-dimensional interacting
Brownian particles model. The total particle number is chosen to be
100. The interacting potential is the spring potential, with a strength
of $\kappa=0.01$. The stiffness of the external harmonic potential
field is $k=0.1$. The initial distribution is such that every particle
is in one of the four spatial regions divided artificially. (b) The
comparison between the true EPR (EPR, the dashed
black line) obtained from an approximate analytical expression, and
our lower bound (the red line).}}

\label{2}
\end{figure}

\normalcolor{We use a system consisting of many interacting Brownian
particles under an external harmonic field (which may be produced
using an optical trap) as an example to illustrate the power of our
bound {[}Fig. \ref{2} (a){]}. This example illustrates its applicability in molecular dynamics simulations and highlights its potential for practical experimental applications.
Even if we know the details of the dynamics, we still need much more
data to measure the EPR (initial positions of
each particle) without our method (see \cite{SupplementalMaterial}
for its approximate analytical expression). In contrast, our
method only requires very coarse-grained data to provide an estimation,
with no prior knowledge of the model details. We define a coarse-grained
state as the state where a randomly picked particle from the system
is in a given spatial region. Then the coarse-grained distribution
$\mathcal{P}(t)$ becomes the spatial distribution of particle number
density in different regions of the space, which is experimentally feasible. For instance,
if the total space is divided into two regions A and B, $\mathcal{P}(t)=(\langle n_{A}\rangle/n,\langle n_{B}\rangle/n)$
where $\langle n_{i}\rangle$ ($i=A,B$) is the average particle number in region $i$ and
$n$ is the total particle number. We remark that the novel coarse-grained mapping employed here differs from the conventional many-to-one mappings in previous literature. We refer to this experimentally beneficial mapping as random coarse-graining (see \cite{bao2025measuring} for more details). The true EPR
and the lower bound are shown in Fig. \ref{2} (b). Our bound could
reproduce over 20\% of the real EPR, which is
significant considering that the data we use to obtain the bound is
very coarse-grained (4 coarse-grained states, compared to the state
space spanned by 100 Brownian particles in continuous space). As the
degree of coarse-graining decreases, the lower bound will be closer
to the real value (see \cite{SupplementalMaterial} for cases with
more coarse-grained states), according to the inverse scaling law
by Yu and Tu \cite{Tu2022scaling}. Note that this example lacks measurable transitions, currents, or trajectories, which precludes the application of any existing thermodynamic inference methods.}  A more complex example
of interacting active particles systems is shown in \cite{SupplementalMaterial}.

\section{Discussion}In this work, we propose a general lower
bound for EP related to the LS constant and KL divergence.
We utilize it to identify a trade-off between intrinsic timescale
and EPR in thermal relaxation. \normalcolor{A
consequence of this trade-off is an inverse speed limit for transforming
states during thermal relaxation, providing a thermodynamic upper
bound for the transformation time. It indicates that a system is unlikely
to remain in a metastable state beyond a certain time threshold. The
inverse speed limit can be extended to open quantum systems, where
an additional contribution from decoherence emerges. }

Our trade-off relation strikingly holds for arbitrary
coarse-grained dynamics, allowing it to be applied to estimate entropy
production irrespective of model details in molecular dynamics and
real experiments. It remains effective even when there are only two
coarse-grained states in relaxation dynamics and no observable current,
which stands in sharp contrast to previous results. Our current study opens up several avenues for future research. Interestingly, similar trade-off relations exist for steady state dissipation rate, a counterintuitive finding that warrants further exploration in a forthcoming paper. Another promising direction would be to investigate these trade-off relations in nonlinear chemical reaction networks. Additionally, while our focus has been on the dominant timescale, it would be valuable to analyze the role of other timescales, aside from
the dominant one.

R. B. is grateful to Guangyi Zou, Tan
Van Vu and Shiling Liang for useful discussions. Part of this work
was finished by R. B. based on the discussion during his participation
in the long-term workshop ``Frontiers in Non-equilibrium Physics 2024
Workshop'' (YITP-T-24-01). R. B. would like to acknowledge the warm
hospitality during his stay in the YITP. This work is supported by
MOST(2022YFA1303100), NSFC (32090044).

R. B. conceived the research, performed the mathematical derivations, designed the numerical simulations and wrote the paper. C. D. performed the numerical simulations. Z. C. contributed to the numerical simulations.
Z. H. supervised the research.

\bibliographystyle{apsrev4-2}
\bibliography{bibfile}

\clearpage

\appendix       

\section{Detailed derivation of Eq. (2) in the main text}

Without loss of generality, we assume transition rates $k_{ij}$ satisfy
the normalization condition $\sum_{i}k_{ij}=1$ throughout this section (here, we temporarily let $k_{ii}$ be the escape rate from $i$, in contrast with the main text).
Releasing the constraint, the only difference is a multiplicative
factor $\sum_{i}k_{ij}$, which will not affect the derivations here.
Under the condition, the stochastic matrix $\mathcal{L}(t)$ in the
main text can be written as $K(t)-\mathbb{I}$, where $K_{ij}(t)=k_{ij}(t)$
and $\mathbb{I}$ is the identity matrix. Then for any function $f$,
the operator $K$ satisfies $(Kf)_{i}=\sum_{j}k_{ij}f_{j}$. Recall
that the the inner product induced by the stationary distribution
is defined as 
\begin{equation}
\langle f,g\rangle_{\pi}=\sum_{i}f_{i}g_{i}^{\dagger}\pi_{i}.
\end{equation}
Based on this inner product, one can further define an adjoint operator
$\mathcal{L}^{\star}$ of $\mathcal{L}$ as $\langle f,\mathcal{L}g\rangle_{\pi}=\langle\mathcal{L}^{\star}f,g\rangle_{\pi}$
for any function $f$ and $g$. Likewise, another adjoint operator
$K^{\star}$ of $K$ is given by $\langle f,Kg\rangle_{\pi}=\langle K^{\star}f,g\rangle_{\pi}$.
Consequently, one can readily check that the operator $K^{\star}$
satisfies $(K^{\star}f)_{j}=\sum_{i}k_{ij}f_{i}$. These relations
will be useful in the derivations below. For more details, see Ref.
\cite{1996logarithmic}. Moreover, one can define the LS constant
with respect to $\mathcal{L}_{s}\equiv(\mathcal{L}+\mathcal{L}^{\star})/2$,
the symmetrized version of $\mathcal{L},$ as
\begin{align}
\lambda_{\text{LS}}= & \inf_{\text{Ent}(f)\neq0}\frac{\text{Re}\langle-\mathcal{L}f,f\rangle_{\pi}}{\text{Ent}(f)}\nonumber \\
= & \inf_{\text{Ent}(f)\neq0}\frac{\langle-\mathcal{L}_{s}f,f\rangle_{\pi}}{\text{Ent}(f)}.
\end{align}
Note that $\mathcal{L}^{\star}=\mathcal{L}$ so that $\mathcal{L}_{s}=\mathcal{L}$
when the detailed balance condition holds. We drop $t$ dependence
in the following for notation's brevity.

\paragraph*{Lemma 1: 
\begin{equation}
\text{Re}\langle-\mathcal{L}^{\star}f,f\rangle_{\pi}=\frac{1}{2}\sum_{i,j}\vert f_{i}-f_{j}\vert^{2}k_{ij}\pi_{j},\label{Lemma1}
\end{equation}
}

where $\vert A\vert\equiv\sqrt{AA^{\dagger}}$.

\paragraph*{Proof: }

Notice that $\text{Re}\langle-\mathcal{L}^{\star}f,f\rangle_{\pi}=\text{Re}\langle(\mathbb{I}-K^{\star})f,f\rangle_{\pi}=\langle f,f\rangle_{\pi}-\text{Re}\langle K^{\star}f,f\rangle_{\pi}$,
and the right hand side of Eq. (\ref{Lemma1}) can be rewritten as
\begin{widetext}
\begin{align}
\frac{1}{2}\sum_{i,j}\vert f_{i}-f_{j}\vert^{2}k_{ij}\pi_{j}= & \frac{1}{2}\sum_{i,j}\left[\vert f_{i}\vert^{2}+\vert f_{j}\vert^{2}-2\text{Re}(f_{i}f_{j}^{\dagger})\right]k_{ij}\pi_{j}\\
= & \frac{\sum_{i}\vert f_{i}\vert^{2}\sum_{j}k_{ij}\pi_{j}+\sum_{j}\vert f_{j}\vert^{2}\pi_{j}\sum_{i}k_{ij}}{2}-\text{Re}\sum_{i,j}k_{ij}f_{i}f_{j}^{\dagger}\pi_{j}\\
= & \frac{\sum_{i}\vert f_{i}\vert^{2}\pi_{i}+\sum_{j}\vert f_{j}\vert^{2}\pi_{j}}{2}-\text{Re}\sum_{j}(K^{\star}f)_{j}f_{j}^{\dagger}\pi_{j}\\
= & \langle f,f\rangle_{\pi}-\text{Re}\langle K^{\star}f,f\rangle_{\pi},
\end{align}
where in the third line, the identities $\sum_{j}k_{ij}\pi_{j}=\pi_{i}$
and $\sum_{i}k_{ij}=1$ have been used. Therefore, $\text{Re}\langle-\mathcal{L}^{\star}f,f\rangle_{\pi}=\langle f,f\rangle_{\pi}-\text{Re}\langle K^{\star}f,f\rangle_{\pi}=\frac{1}{2}\sum_{i,j}\vert f_{i}-f_{j}\vert^{2}k_{ij}\pi_{j}$.

Further, with the detailed balance condition $k_{ij}\pi_{j}=k_{ji}\pi_{i}$
holding, one can similarly show that
\begin{equation}
\text{Re}\langle-\mathcal{L}f,g\rangle_{\pi}=\frac{1}{2}\sum_{i,j}\left(f_{i}-f_{j}\right)\left(g_{i}-g_{j}\right)k_{ij}\pi_{j}.\label{Lemma1b}
\end{equation}
Note that $f$ can be a complex function and $f_{i}^{\dagger}$ denote
the complex conjugate of $f_{i}$. 

\paragraph*{Lemma 2: }

For a system with detailed balance condition ($\boldsymbol{\pi}=\boldsymbol{p}^{\text{eq}}$),
any function $f$ in the state space of the system satisfies: 
\begin{equation}
\langle-\mathcal{L}^{\star}\ln f,f\rangle_{\pi}\geq4\langle-\mathcal{L}^{\star}\sqrt{f},\sqrt{f}\rangle_{\pi}.
\end{equation}
Additionally, in the absence of detailed balance condition, a weaker
inequality 
\begin{equation}
\langle-\mathcal{L}^{\star}\ln f,f\rangle_{\pi}\geq2\langle-\mathcal{L}^{\star}\sqrt{f},\sqrt{f}\rangle_{\pi}
\end{equation}
holds.

\paragraph*{Proof: }

For any $a,b>0$, 
\begin{align}
\left(\frac{\sqrt{a}-\sqrt{b}}{a-b}\right)^{2}= & \left[\frac{1}{2(a-b)}\int_{b}^{a}x^{-\frac{1}{2}}\text{d}x\right]^{2}\nonumber \\
\leq & \left[\int_{b}^{a}\frac{1}{4(a-b)^{2}}\text{d}x\right]\left(\int_{b}^{a}x^{-1}\text{d}x\right)=\frac{1}{4}\frac{\ln a-\ln b}{a-b}
\end{align}
thus the inequality below is fulfilled:
\begin{align}
\left(a-b\right)\left[\ln a-\ln b\right] & \geq4\left(\sqrt{a}-\sqrt{b}\right)^{2}.
\end{align}
Then using Lemma 1 and Eq. (\ref{Lemma1b}), the inequality $\langle-\mathcal{L}^{\star}\ln f,f\rangle_{\pi}\geq4\langle-\mathcal{L}^{\star}\sqrt{f},\sqrt{f}\rangle_{\pi}$
is immediately derived. For any $a,b>0$, there is another inequality
\begin{equation}
\ln a^{2}-\ln b^{2}\leq\frac{2(a-b)}{b}
\end{equation}
due to the concavity of the function $\ln x^{2}$. Multiplying both
sides by $b^{2}$ leads to 
\begin{equation}
b^{2}(\ln a^{2}-\ln b^{2})\leq2b(a-b).
\end{equation}
Then let $f_{i}=b^{2}$ and $(K\ln f)_{i}=\ln a^{2}$, one obtains
\begin{equation}
f_{i}[(K^{\star}-\mathbb{I})\ln f]_{i}\leq2\sqrt{f_{i}}(\sqrt{e^{K^{\star}\ln f}}-\sqrt{f})_{i}\leq2\sqrt{f_{i}}[(K^{\star}-\mathbb{I})\sqrt{f}]_{i},\label{neq}
\end{equation}
where the inequality
\begin{eqnarray*}
(\sqrt{e^{K^{\star}\ln f}})_{i}=\sqrt{e^{\sum_{j}k_{ji}(\ln f)_{j}}} & \leq & \sum_{j}k_{ji}(\sqrt{e^{\ln f_{j}}})=K^{\star}\sqrt{f}
\end{eqnarray*}
has been used (the inequality is from the convexity of the function
$\sqrt{e^{x}}$ and the Jensen inequality). Notice that $-\mathcal{L}^{\star}=\mathbb{I}-K^{\star}$,
thus Eq. (\ref{neq}) is equal to
\begin{equation}
[(-\mathcal{L}^{\star})\ln f]_{i}f_{i}\pi_{i}\geq2[(-\mathcal{L}^{\star})\sqrt{f}]_{i}\sqrt{f_{i}}\pi_{i},
\end{equation}
which directly yields $\langle-\mathcal{L}^{\star}\ln f,f\rangle_{\pi}\geq2\langle-\mathcal{L}^{\star}\sqrt{f},\sqrt{f}\rangle_{\pi}$.

\paragraph*{Proof of the Eq. (2) in the main text: }

Equipped with Lemma 1 and Lemma 2, we can prove the inequality (3)
(with detailed balance) as follow:

\begin{align}
\frac{\text{d}}{\text{d}t}D[\boldsymbol{p}(t)\vert\vert\boldsymbol{\pi}_{\tau}]|_{\tau=t}= & \sum_{i}[1+\ln\frac{p_{i}(t)}{\pi_{i}^{t}}]\dot{p}_{i}(t)\\
= & \sum_{i}\ln\frac{p_{i}(t)}{\pi_{i}^{t}}\frac{\text{d}}{\text{d}t}\left[\frac{p_{i}(t)}{\pi_{i}^{t}}\right]\pi_{i}\\
= & -\sum_{i}\ln\frac{p_{i}(t)}{\pi_{i}^{t}}\left[-\mathcal{L}\frac{\boldsymbol{p}(t)}{\boldsymbol{\pi}_{t}}\right]_{i}\pi_{i}^{t}\\
= & -\langle\ln\frac{\boldsymbol{p}(t)}{\boldsymbol{\pi}_{t}},-\mathcal{L}\frac{\boldsymbol{p}(t)}{\boldsymbol{\pi}_{t}}\rangle_{\pi}\\
\equiv & -\langle\ln f(t),-\mathcal{L}f(t)\rangle_{\pi}\\
= & -\langle-\mathcal{L}^{\star}\ln f(t),f(t)\rangle_{\pi}\\
\leq & -4\langle-\mathcal{L}^{\star}\sqrt{f(t)},\sqrt{f(t)}\rangle_{\pi}\label{DB}\\
= & -4\text{Re}\langle-\mathcal{L}\sqrt{f(t)},\sqrt{f(t)}\rangle_{\pi}\leq-4\lambda_{\text{LS}}(t)\text{Ent}(\sqrt{f(t)})\\
= & -4\lambda_{\text{LS}}(t)\sum_{i=1}^{N}\vert\sqrt{\frac{p_{i}(t)}{\pi_{i}^{t}}}\vert^{2}\ln\left(\frac{\vert\sqrt{\frac{p_{i}(t)}{\pi_{i}}}\vert^{2}}{\langle\sqrt{\frac{\boldsymbol{p}(t)}{\boldsymbol{\pi}_{t}}},\sqrt{\frac{\boldsymbol{p}(t)}{\boldsymbol{\pi}_{t}}}\rangle_{\pi}}\right)\pi_{i}^{t}\\
= & -4\lambda_{\text{LS}}(t)D[\boldsymbol{p}(t)\vert\vert\boldsymbol{\pi}_{t}].
\end{align}
\end{widetext}

It should be noted that, since $f(t)=\frac{\boldsymbol{p}(t)}{\boldsymbol{\pi}_{t}}$
is a real function and $\langle-\mathcal{L}^{\star}f,f\rangle_{\pi}=\langle f,-\mathcal{L}f\rangle_{\pi}=\langle-\mathcal{L}f,f\rangle_{\pi}^{\dagger}$,
we get that $\langle-\mathcal{L}^{\star}f,f\rangle_{\pi}=\text{Re}\langle-\mathcal{L}^{\star}f,f\rangle_{\pi}=\text{Re}\langle-\mathcal{L}f,f\rangle_{\pi}$,
which has been used in the third last line. Without detailed balance,
Eq. (\ref{DB}) should be substituted with $-2\langle-\mathcal{L}^{\star}\sqrt{f(t)},\sqrt{f(t)}\rangle_{\pi}$,
where the only difference is a multiplicative constant $1/2$.

The discussions above can be naturally generalized to the system coupled
to multiple heat baths, in which the stochastic matrix consists of
contributions from each independent baths as $\mathcal{L}(t)=\sum_{\nu}\mathcal{L}^{\nu}(t)$,
with $\mathcal{L}^{\nu}(t)$ being the stochastic matrix related to
the $\nu$th bath. With multiple heat baths, the non-adiabatic entropy
production rate can still be associated with the KL divergence as
\begin{equation}
\dot{\sigma}^{\text{na}}(t)=-\sum_{i}\dot{p}_{i}(t)\ln\frac{p_{i}(t)}{\pi_{i}^{t}}=-\frac{\text{d}}{\text{d}t}D[\boldsymbol{p}(t)\vert\vert\boldsymbol{\pi}_{\tau}]|_{\tau=t}.
\end{equation}
Notably, $\dot{\sigma}^{\text{na}}(t)$ is only a function of the
coarse-grained transition rates $k_{ij}(t)=\sum_{\nu}k_{ij}^{\nu}(t)$,
which is not pertinent to the individual contribution from the $\nu$th
heat bath. Therefore, the general bound (6) in the main text can be
directly generalized to the system coupled to multiple heat baths
as
\begin{equation}
\dot{\sigma}(t)\geq\dot{\sigma}^{\text{na}}(t)\geq2\lambda_{\text{LS}}(t)D[\boldsymbol{p}(t)\vert\vert\boldsymbol{\pi}_{t}].
\end{equation}
 As mentioned in the main text, the total EPR
can be decomposed into two parts, one part is the non-adiabatic entropy
production, and another part (housekeeping or adiabatic EP
rate) reads
\begin{equation}
\dot{\sigma}^{\text{hs}}(t)=\sum_{\nu}\sum_{i,j}k_{ij}^{\nu}(t)p_{j}(t)\ln\frac{k_{ij}^{\nu}(t)\pi_{j}^{t}}{k_{ji}^{\nu}(t)\pi_{i}^{t}}.
\end{equation}
It can be seen from this expression that only if the transitions induced
by every heat bath all satisfy the detailed balance condition, i.e.,
$k_{ij}^{\nu}(t)\pi_{j}^{t}=k_{ji}^{\nu}(t)\pi_{i}^{t}$ for any $\nu$,
will the housekeeping part vanish (so that $\boldsymbol{\pi}_{t}=\boldsymbol{p}_{t}^{\text{eq}}$).
In this case, $\dot{\sigma}(t)=\dot{\sigma}^{\text{na}}(t)\geq4\lambda_{\text{LS}}(t)D[\boldsymbol{p}(t)\vert\vert\boldsymbol{p}_{t}^{\text{eq}}]$.

\section{An application of Eq. (3) to time-dependent dynamics}

In this section, we are interested in an example considered in Refs.
\cite{busiello2021dissipation,23PRE_Lu}, where the transition matrix
is under periodic oscillations. This setting has actual applications
in biological and chemical systems. As a result, the dynamics is governed
by a time-dependent stochastic matrix : $\frac{\text{d}}{\text{d}t}\boldsymbol{p}(t)=\mathcal{L}(t)\boldsymbol{p}(t),$
where
\begin{equation}
\mathcal{L}(t)=\begin{cases}
\mathcal{L}_{1} & t\in[2n\tau,(2n+1)\tau]\\
\mathcal{L}_{2} & t\in[(2n+1)\tau,(2n+2)\tau]
\end{cases},\ n\in\mathbb{N}.
\end{equation}
Here, $\mathcal{L}_{1}$ and $\mathcal{L}_{2}$ are time-independent
stochastic matrices satisfying the detailed balance condition and
$\tau$ is the period of oscillation. The LS constant of $\mathcal{L}_{1}$
and $\mathcal{L}_{2}$ are denoted as $\lambda_{\text{LS,1}}$ and
$\lambda_{\text{LS,2}}$ respectively. The system under this setting
will finally converge to a periodic stationary state in which $\boldsymbol{p}(t)=\boldsymbol{p}(t+\tau)$.

In the fast oscillation limit $\tau\rightarrow0$, it has been demonstrated
that the periodic stationary state reduces to an effective equilibrium
state $\boldsymbol{p}^{\text{eff}}$ corresponding to an effective
stochastic matrix 
\begin{equation}
\mathcal{L}^{\text{eff}}\equiv\frac{\mathcal{L}_{1}+\mathcal{L}_{2}}{2},
\end{equation}
i.e., 
\begin{equation}
\mathcal{L}^{\text{eff}}\boldsymbol{p}^{\text{eff}}=0.
\end{equation}
The effective LS constant corresponding to the effective stochastic
matrix is given by 
\begin{align}
\lambda_{\text{LS}}^{\text{eff}}= & \inf_{\text{Ent}(f)\neq0}\frac{\text{Re}\langle-\mathcal{\mathcal{L}^{\text{eff}}}f,f\rangle_{\pi}}{\text{Ent}(f)}\nonumber \\
= & \frac{1}{2}\inf_{\text{Ent}(f)\neq0}\frac{\text{Re}\langle-\left(\mathcal{L}_{1}+\mathcal{L}_{2}\right)f,f\rangle_{\pi}}{\text{Ent}(f)}\nonumber \\
= & \frac{\inf_{\text{Ent}(f)\neq0}\frac{\text{Re}\langle-\mathcal{L}_{1}f,f\rangle_{\pi}}{\text{Ent}(f)}+\inf_{\text{Ent}(f)\neq0}\frac{\text{Re}\langle-\mathcal{L}_{2}f,f\rangle_{\pi}}{\text{Ent}(f)}}{2}\nonumber \\
= & \frac{\lambda_{\text{LS,1}}+\lambda_{\text{LS,2}}}{2}.
\end{align}
Then, applying our first main result Eq. (7) to the situation when
the system has reached the effective equilibrium yields that 
\begin{equation}
\text{\ensuremath{\dot{\sigma}(t)}}\geq2\left(\lambda_{\text{LS,1}}+\lambda_{\text{LS,2}}\right)D\left[\boldsymbol{p}^{\text{eff}}\vert\vert\boldsymbol{p}_{t}^{\text{eq}}\right],\label{EffIne}
\end{equation}
since the system is in the effective equilibrium state and the effective
LS constant is given by $(\lambda_{\text{LS,1}}+\lambda_{\text{LS,2}})/2$.
Note that the instantaneous equilibrium distribution $\boldsymbol{p}_{t}^{\text{eq}}$
at time $t$ will be one of the equilibrium distributions $\boldsymbol{p}_{1}^{\text{eq}}$
or $\boldsymbol{p}_{2}^{\text{eq}}$ corresponding to $\mathcal{L}_{1}$
or $\mathcal{L}_{2}$, which are not matched with the effective equilibrium
state. As a consequence, $D\left[\boldsymbol{p}^{\text{eff}}\vert\vert\boldsymbol{p}_{t}^{\text{eq}}\right]>0$
so that the lower bound given by Eq. (\ref{EffIne}) is positive.
For a very large time $t\gg\tau$, one can further bound the entropy
production $\sigma_{[0,t]}$ during the interval $[0,t]$ asymptotically
from below as 
\begin{equation}
\sigma_{[0,t]}\gtrsim\left(2\lambda_{\text{LS,1}}D\left[\boldsymbol{p}^{\text{eff}}\vert\vert\boldsymbol{p}_{1}^{\text{eq}}\right]+2\lambda_{\text{LS,2}}D\left[\boldsymbol{p}^{\text{eff}}\vert\vert\boldsymbol{p}_{2}^{\text{eq}}\right]\right)t
\end{equation}
This is interesting because the system in an effective equilibrium
state may not be distinguished from a real equilibrium state in a
coarse grained level, e.g., in the experimental observations level,
which may lead to the wrong conclusion that there is no EP.
However, our positive bound can recover at least part of the ``hidden''
EP (rate), which is notably stronger than the conventional
second law of thermodynamics. We should emphasize that even when the
system is not periodically switching, but randomly switching between
two configurations $\mathcal{L}_{1}$ and $\mathcal{L}_{2}$ at a
constant Poisson rate $r$, the above result can still apply in the
fast switching limit $r\rightarrow\infty$, when the system is still
in an effective equilibrium state corresponding to $\mathcal{L}^{\text{eff}}$.

Our bound is not limited to the fast oscillation limit, when the period
$\tau$ is finite, our lower bound can still be applied and probably
give a positive value. For example, assuming $t>\tau$ and $\mathcal{L}(0)=\mathcal{L}_{1}$,
one can utilize Eq. (6) to obtain a positive lower bound for $\sigma_{[0,t]}$
in this case: 
\begin{widetext}

\begin{equation}
\sigma_{[0,t]}\geq\left[1-e^{-4\lambda_{\text{LS,1}}\tau}\right]D\left[\boldsymbol{p}(0)\vert\vert\boldsymbol{p}_{\mathcal{L}_{1}}^{\text{eq}}\right]+\left[1-e^{-4\lambda_{\text{LS,2}}(t-\tau)}\right]D\left[\boldsymbol{p}(\tau)\vert\vert\boldsymbol{p}_{\mathcal{L}_{2}}^{\text{eq}}\right],
\end{equation}
where $\boldsymbol{p}(\tau)=\mathcal{T}e^{\int_{0}^{\tau}\mathcal{L}_{1}\text{d}t}\boldsymbol{p}(0)$,
$\mathcal{T}$ is the time-ordering.
    
\end{widetext}
\section{Properties of the LS constant and its relation to the spectral gap}

The LS constant characterizes the intrinsic relaxation timescale (mixing
time) as \cite{1996logarithmic}
\begin{equation}
    \frac{1}{2\lambda_{\text{LS}}}\leq\tau_{\text{rel}}\leq\frac{4+\log\log[1/\pi_{\star}]}{2\lambda_{\text{LS}}},
\end{equation}
where $\pi_{\star}\equiv\min_{i}\pi_{i}$ and the relaxation time
$\tau_{\text{rel}}$ is defined as
\begin{equation}
    \tau_{\text{rel}}=\inf\left\{ t>0:\sup_{i}\left\Vert \frac{\boldsymbol{p}(t|p_{j}(0)=\delta_{ij})}{\boldsymbol{\pi}}-\mathbb{I}\right\Vert _{2}\leq\frac{1}{e}\right\},
\end{equation}
with the $L_{2}$ norm being defined as $||f||_{2}\equiv\sqrt{\langle f,f\rangle_{\pi}}=\sqrt{\sum_{i}|f_{i}|^{2}\pi_{i}}$
and $\mathbb{I}$ being the unit vector. \normalcolor{Another
definition of the relaxation time is given by }

\normalcolor{
\begin{equation}
\tau_{\text{rel}}\equiv\frac{1}{2\lambda_{\text{LS}}}=-\lim_{t\rightarrow\infty}\frac{t}{2\ln D\left[\boldsymbol{p}(t)\vert\vert\boldsymbol{p}^{\text{eq}}\right]},
\end{equation}
which seems require more statistical data to produce a precise value.
The advantage of the above definition of $\tau_{\text{rel}}$ is that
it can be applied to any non-Markovian coarse-grained dynamics without
assuming timescale separation. By contrast, the initial definition
$\tau_{\text{rel}}=\inf\left\{ t>0:\sup_{i}\left\Vert \frac{\boldsymbol{p}(t|p_{j}(0)=\delta_{ij})}{\boldsymbol{\pi}}-\mathbb{I}\right\Vert _{2}\leq\frac{1}{e}\right\} $
can only be effective for fine-grained Markov dynamics and the coarse-grained
relaxation dynamics whose KL divergence from the equilibrium state
decays exponentially in the large time limit. To assure exponential
decay in the large time limit, a clear timescale separation is required.
Without timescale separation, the maximization over the initial distribution
on coarse-grained level cannot uniquely determine the relaxation timescale
$\tau_{\text{rel}}^{CG}$, because different microscopic distributions
within coarse-grained states will affect $\tau_{\text{rel}}^{CG}$.}

When the unique stationary distribution $\boldsymbol{\pi}$ is an
equilibrium distribution $\boldsymbol{p}^{\text{eq}}$, the upper
bound of $\tau_{\text{rel}}$ can be enhanced by a factor $1/2$.
There are similar inequalities for $\tau_{\text{rel}}$ using the
spectral gap $\lambda_{g}$ when detailed balance condition holds,
i.e., 
\[
\frac{1}{\lambda_{g}}\leq\tau_{\text{rel}}\leq\frac{2+\log[1/\pi_{\star}]}{2\lambda_{g}}.
\]

Further, there is a hierarchical relation between the spectral gap
$\lambda_{g}$ and LS constant $\lambda_{\text{LS}}$ \cite{gine2006lectures}:

\begin{align}
\frac{\lambda_{g}}{2}\geq\lambda_{\text{LS}}&\geq\frac{1-2\pi_{\star}}{\ln[(1-\pi_{\star})/\pi_{\star}]} \nonumber\\
&\lambda_{g}\geq\max\left\{ \frac{1-2\pi_{i}}{\ln[(1-\pi_{i})/\pi_{i}]}\lambda_{g},0\right\} .
\end{align}
The spectral gap $\lambda_{g}$ is the second largest eigenvalue of
$-\mathcal{L}_{s}$, and it has a similar definition to $\lambda_{\text{LS}}$
as
\begin{align}
\lambda_{g}= & \inf_{\langle f,f\rangle_{\pi}\neq0}\frac{\text{Re}\langle-\mathcal{\mathcal{L}}f,f\rangle_{\pi}}{\langle f,f\rangle_{\pi}}\nonumber \\
= & \inf_{\langle f,f\rangle_{\pi}\neq0}\frac{\langle-\mathcal{\mathcal{L}}_{s}f,f\rangle_{\pi}}{\langle f,f\rangle_{\pi}}.
\end{align}
Note that when the detailed balance condition holds, $\mathcal{\mathcal{L}}_{s}=\mathcal{\mathcal{L}}$
so that $\lambda_{g}$ becomes the second largest eigenvalue of $-\mathcal{L}$
in this case. 

Consequently, $\lambda_{\text{LS}}$ may characterize the relaxation
timescale better compared with the spectral gap $\lambda_{g}$ due
to the hierarchical relation above (the inequality from $\lambda_{\text{LS}}$
is tighter than the inequality from $\lambda_{g}$).

Due to the close connection between $\lambda_{\text{LS}}$ and the
spectral gap $\lambda_{g}$ which is usually easier to determined,
one can obtain another useful bound related to $\lambda_{g}$ as
\begin{equation}
\dot{\sigma}(t)\geq4C\lambda_{g}\sigma_{\text{tot}}^{t},\label{bound2}
\end{equation}
where $C=(1-2\pi_{\star})/\ln(1-\pi_{\star}/\pi_{\star})$. This bound
uncovers a connection between the thermodynamic irreversibility and
the spectrum of the dynamical generator in thermal relaxation.

\section{\normalcolor{Theoretical justifications of the coarse-grained
results (11)}}

\subsection{\textit{\normalcolor{The case when there is time-scale separation
and the coarse-graining is appropriate}}}

In this case, we can rigorously prove that the trade-off relation
for coarse-grained dynamics, i.e., the Eq. (11) in the main text is
valid by using the data-processing inequality (or log-sum inequality).\normalcolor{{}
The data-processing inequality reads}

\normalcolor{
\begin{equation}
D[p(x)||q(x)]\geq D[p(y)||q(y)],
\end{equation}
where $y=f(x)$ is an arbitrary function of $x$. This follows from
the chain rule of KL divergence, i.e., 
\begin{align}
D[p(x,y)||q(x,y)]= & D[p(y|x)||q(y|x)]+D[p(x)||q(x)]\nonumber \\
= & D[p(x|y)||q(x|y)]+D[p(y)||q(y)].
\end{align}
The term $D[p(y|x)||q(y|x)]=0$ because $p(y|x)=q(y|x)=1$ only when
$y=f(x)$ and $p(y|x)=q(y|x)=0$ otherwise. $D[p(x|y)||q(x|y)]\geq0$
and the equality holds when $y=f(x)$ is a one-to-one mapping. Thus,
$D[p(x)||q(x)]\geq D[p(y)||q(y)]$. Choosing $y=\sum_{i}c_{i}\mathbb{I}_{A_{i}}$,
where the universal set $\chi=A_{1}+A_{2}+...+A_{n}$, leads to 
\begin{equation}
D[p(x)||q(x)]\geq\sum_{i}p(x\in A_{i})\ln\frac{p(x\in A_{i})}{q(x\in A_{i})}\equiv D[\mathcal{P}||\mathcal{Q}],
\end{equation}
where $\mathcal{P}_{i}\equiv p(x\in A_{i})$ is the probability that
the system is in coarse-grained state $i$. Due to the arbitrariness
of the set $A_{i}$, this prove that the KL divergence between always
decreases under any conceivable state coarse-graining. Therefore,
$D\left[\boldsymbol{p}(t)\vert\vert\boldsymbol{p}^{\text{eq}}\right]\geq D[\mathcal{P}(t)\vert\vert\mathcal{P}^{eq}]$
and 
\begin{equation}
\dot{\sigma}(t)\geq\frac{2}{\tau_{\text{rel}}}D\left[\boldsymbol{p}(t)\vert\vert\boldsymbol{p}^{\text{eq}}\right]\geq\frac{2}{\tau_{\text{rel}}^{CG}}D[\mathcal{P}(t)\vert\vert\mathcal{P}^{eq}]
\end{equation}
whenever $\tau_{\text{rel}}\sim\tau_{\text{rel}}^{CG}$ (when $\frac{\tau_{\text{rel}}^{CG}}{\tau_{\text{rel}}}\geq\frac{D[\mathcal{P}(t)\vert\vert\mathcal{P}^{eq}]}{D\left[\boldsymbol{p}(t)\vert\vert\boldsymbol{p}^{\text{eq}}\right]}$),
which is a criterion of good coarse-graining mapping. In other words,
the breakdown of the above inequality is a strong witness of inappropriate
coarse-graining procedure (but we argue that this equality holds for
general coarse-graining). }

\subsection{\textit{\normalcolor{General coarse-grained dynamics }}}

\normalcolor{In general, we only know that the probability distribution
of microscopic dynamics evolves according to the master equation $\frac{d}{dt}p_{i}(t)=\sum_{ij}k_{ij}(t)p_{j}-k_{ji}(t)p_{i}$
and each coarse-grained state consists of many microscopic states.
In this setting, the coarse-grained dynamics can be described by an
effective master equation,
\begin{equation}
\frac{d}{dt}\mathcal{P}_{m}(t)=\sum_{n}k_{mn}^{CG}(t)\mathcal{P}_{n}-k_{nm}^{CG}(t)\mathcal{P}_{m},
\end{equation}
where the time-dependent coarse-grained transition rate $k_{mn}^{CG}(t)$
reads
\begin{equation}
k_{mn}^{CG}(t)=\sum_{i\in m}\sum_{j\in n}k_{ij}p_{t}(j|n),
\end{equation}
where $p_{t}(j|n)=\frac{p_{j}(t)}{\sum_{x\in n}p_{x}(t)}$ is the
conditional probability of the system being in microscopic state $j$
at time $t$ given that it is in the coarse-grained state $n$. The
dynamics under this effective master equation is in general non-Markovian
because transition rates here are dependent on the probability distribution
of microscopic states.}

\normalcolor{If the detailed balance condition $k_{ij}(t)p_{j}^{eq}=k_{ji}(t)p_{i}^{eq}$
holds in the fine-grained level, in the coarse-grained level we still
have that $k_{mn}^{CG}(t)\mathcal{P}_{n}^{eq}=k_{nm}^{CG}(t)\mathcal{P}_{m}^{eq}$
for any pair of coarse-grained states $m,n$. If this is not the case,
the coarse-grained dynamics can be nonequilibrium even when the original
fine-grained dynamics is in equilibrium, which is not physical. With
detailed balance condition, one can write that
\begin{align}
& -\frac{\text{d}}{\text{d}t}D[\mathcal{P}(t)\vert\vert\mathcal{P}^{eq}]|_{\tau=t} =  \sum_{i}\dot{\mathcal{P}}_{i}(t)\ln\frac{\mathcal{P}_{i}^{eq}}{\mathcal{P}_{i}(t)} \nonumber\\
& =  \sum_{i,j}k_{ij}^{CG}(t)\mathcal{P}_{j}(t)\ln\frac{k_{ij}^{CG}(t)\mathcal{P}_{j}(t)}{k_{ji}^{CG}(t)\mathcal{P}_{i}(t)}\nonumber\\
& =  \mathcal{K}_{CG}\sum_{l_{CG}}p(l_{CG})\ln\frac{p(l_{CG})}{p(\tilde{l}_{CG})}\nonumber\\
& \leq  \mathcal{K}\sum_{l}p(l)\ln\frac{p(l)}{p(\tilde{l})}\nonumber\\
& =  \sum_{i,j}k_{ij}(t)p_{j}(t)\ln\frac{k_{ij}(t)p_{j}(t)}{k_{ji}(t)p_{i}(t)}=\dot{\sigma}(t)
\end{align}
Here, $\mathcal{K}_{CG}$ and $\mathcal{K}$ are dynamical activities
in the coarse-grained and fine-grained level, quantifying the mean
time between two consecutive jumps in different level. $l$, $\tilde{l}$
denote microscopic transitions and their time-reversal transitions,
and $l_{CG},\tilde{l}_{CG}$ are their coarse-grained counterparts.
In the second line, the detailed balance condition has been used.
The last inequality is due to data-processing inequality. $\lambda_{LS}^{CG}(t)$
for the coarse-grained dynamics can be defined using the time-dependent
generator $\mathcal{L}^{CG}$. In addition, the inequality 
\begin{equation}
-\frac{\text{d}}{\text{d}t}D[\mathcal{P}(t)\vert\vert\mathcal{P}^{eq}]|_{\tau=t}\geq4\lambda_{LS}^{CG}(t)D[\mathcal{P}(t)\vert\vert\mathcal{P}^{eq}]
\end{equation}
is valid, because the proof of the inequality only relies on the mathematical
form of the master equation. $\lambda_{LS}^{CG}(t)$ still quantifies
the largest relaxation timescale at time $t$. Combining the above
two inequalities, we deduce that the inequality
\begin{equation}
\dot{\sigma}(t)\geq4\lambda_{LS}^{CG}(t)D[\mathcal{P}(t)\vert\vert\mathcal{P}^{eq}]\label{CGLB}
\end{equation}
hold for coarse-grained relaxation dynamics. Eq. (\ref{CGLB}) is
a stronger lower bound than the Eq. (12) in the main text. Define
$\lambda_{LS}^{CG}\equiv\min_{t}\{\lambda_{LS}^{CG}(t)\}$, we then
conclude that $\tau_{\text{rel}}^{\text{CG}}\equiv\inf_{\tau}\left\{ \sup_{i}\left\Vert \frac{\mathcal{P}(\tau|\mathcal{P}_{j}(0)=\delta_{ij})}{\mathcal{P}^{\text{eq}}}-\mathbb{I}\right\Vert _{2}\leq1/e\right\} $
(the definition works only when there is a timescale separation in
fine-grained dynamics) is lower bounded as 
\begin{equation}
\tau_{\text{rel}}^{\text{CG}}\geq\frac{1}{2\lambda_{LS}^{CG}},
\end{equation}
so that we can replace $\lambda_{LS}^{CG}(t)$ with $\frac{2}{\tau_{\text{rel}}^{\text{CG}}}$
for every $t$ in Eq. (\ref{CGLB}). }

\normalcolor{For another definition of the coarse-grained relaxation
timescale, $\tau_{\text{rel}}^{CG}\equiv\lim_{t\rightarrow\infty}\frac{-t}{2\ln D\left[\mathcal{P}(t)\vert\vert\mathcal{P}^{\text{eq}}\right]}$,
we argue that it satisfies $\frac{1}{2\lambda_{LS}^{CG}(t)}\leq\tau_{\text{rel}}^{CG}\leq\tau_{\text{rel}}$
for any $t$. By definition, $\tau_{\text{rel}}^{CG}=\lim_{t\rightarrow\infty}\frac{-t}{2\ln D\left[\mathcal{P}(t)\vert\vert\mathcal{P}^{\text{eq}}\right]}\leq\tau_{\text{rel}}\equiv\lim_{t\rightarrow\infty}\frac{-t}{2\ln D\left[\boldsymbol{p}(t)\vert\vert\boldsymbol{p}^{\text{eq}}\right]}$,
due to the data-processing inequality $D\left[\mathcal{P}(t)\vert\vert\mathcal{P}^{\text{eq}}\right]\leq D\left[\boldsymbol{p}(t)\vert\vert\boldsymbol{p}^{\text{eq}}\right]$
as shown in the part A of this section. The first inequality also
holds with physical assumptions that }

\normalcolor{1. $\lim_{t\rightarrow\infty}\frac{1}{2\lambda_{LS}^{CG}(t)}$
exists and converges to $\tau_{\text{rel}}^{CG}$. }

\normalcolor{2. $\min_{t}\{\lambda_{LS}^{CG}(t)\}=\lim_{t\rightarrow\infty}\lambda_{LS}^{CG}(t)$. }

\normalcolor{Although the assumptions cannot be rigorously proven,
the physics that all faster relaxation modes vanish in the large time
limit strongly supports their validity. Consequently, we obtain the
desired lower bound, Eq. (12) in the main text, which is experimentally
feasible. Note that when the detailed balance condition is broken,
i.e., when the stationary state is not an equilibrium state, the justification
above does not hold because the data-processing inequality cannot
be applied. However, if the coarse-graining mapping is appropriate
(as shown in part A), the coarse-grained lower bound for the non-adiabatic
EPR still holds.}

\normalcolor{Additionally, it should be noted that the Markov jump
process can be obtained by faithfully discretizing the continuous
Fokker-Planck equations \cite{shiraishi16tradeoff}, which assures
that our trade-off relation holds for Langevin dynamics by taking
the continuum limits. Therefore, our results can be applied to molecular
dynamics simulations, where the underlying dynamics can be described
by Langevin equations.}

\section{\normalcolor{Analytical calculation of the EP
rate for the interacting Brownian particles system}}

Here, we derive an analytical expression for the EPR of a single Brownian particle in a harmonic potential field. We then use this single-particle expression to approximate the EPR of the interacting Brownian particle system discussed in the main text.

\normalcolor{Considering a single Brownian particle under a harmonic
field $U(\boldsymbol{x})=k||\boldsymbol{x}-\boldsymbol{a}||^{2}/2$
in 2-dimensional space, where $k$ is the stiffness and $\boldsymbol{a}=(a_{x},a_{y})$
is the center of the field. The corresponding Langevin function reads
$\boldsymbol{\dot{x}}=-\mu\nabla U(\boldsymbol{x})+\sqrt{2D}\boldsymbol{\xi}(t)$.
The probability distribution of the position $\boldsymbol{x}=(x,y)$
of the particle evolves according to the Fokker-Planck equation
\begin{equation}
\frac{\partial p_{t}(x,y)}{\partial t}=\mu\left[\partial_{x}\left(k(x-a_{x})-T\partial_{x}\right)+\partial_{y}\left(k(y-a_{y})-T\partial_{y}\right)\right]p_{t}(x,y),
\end{equation}
with $\mu$ being the mobility and satisfying $\mu=D/k_{B}T$. Multiplying
both sides with $x$ (or $y$) and then integrating both sides with
respect to $y$ (or $x$), leading to two decoupled equations for
the first moments as 
\begin{align}
d_{t}\langle x\rangle_{t} & =-\mu k\left(\langle x\rangle_{t}-a_{x}\right)\nonumber \\
d_{t}\langle y\rangle_{t} & =-\mu k\left(\langle y\rangle_{t}-a_{y}\right),
\end{align}
where the conservation of probability has been used. We let $a_{x}=a_{y}=0$,
i.e., the harmonic field is posed at the center of the space. Then
the moments at time $t$ are solved as 
\begin{equation}
\langle x\rangle_{t}=\langle x\rangle_{0}e^{-\mu kt},\ \langle y\rangle_{t}=\langle y\rangle_{0}e^{-\mu kt}.
\end{equation}
If the initial distribution is Gaussian, the probability $p_{t}(x,y)$
will keep Gaussian at any time, i.e., 
\[
p_{t}(x,y)=\frac{k}{2\pi T}\exp\left(-\frac{k||\boldsymbol{x}-\langle\boldsymbol{x}\rangle_{t}||^{2}}{2T}\right).
\]
The EPR is 
\begin{equation}
\dot{\sigma}(t)=\frac{1}{\mu T}\int d\boldsymbol{x}||\boldsymbol{\nu}_{t}(\boldsymbol{x})||^{2}p_{t}(\boldsymbol{x}),
\end{equation}
where the local mean velocity is defined as 
\begin{equation}
\boldsymbol{\nu}_{t}(\boldsymbol{x})\equiv\mu\left(-\nabla U(\boldsymbol{x})-T\nabla\ln p_{t}(\boldsymbol{x})\right).
\end{equation}
Using the expressions for $p_{t}(\boldsymbol{x})=p_{t}(x,y)$ and
$\langle x\rangle_{t},\ \langle y\rangle_{t}$, we obtain that the
EPR for a single Brownian particle at time $t$
is 
\begin{equation}
\dot{\sigma}(t)=\frac{Dk^{2}}{k_{B}T^{2}}\left(\langle x\rangle_{0}^{2}+\langle y\rangle_{0}^{2}\right)e^{-\frac{2Dk}{k_{B}T}t}.
\end{equation}
Since the interacting potential in our case is also harmonic, and
the interaction strength $\kappa\ll k$, we conclude that the total
EPR for $N$ such Brownian particles at time $t$
is simply
\begin{equation}
\dot{\sigma}_{tot}(t)\approx\frac{NDk^{2}}{k_{B}T^{2}}\left(\overline{\langle x\rangle_{0}^{2}}+\overline{\langle y\rangle_{0}^{2}}\right)e^{-\frac{2Dk}{k_{B}T}t},
\end{equation}
where $\overline{\langle x\rangle_{0}^{2}}\equiv\frac{1}{N}\sum_{k}\langle x_{k}\rangle_{0}^{2}$
and $\overline{\langle y\rangle_{0}^{2}}\equiv\frac{1}{N}\sum_{k}\langle y_{k}\rangle_{0}^{2}$.
This expression may slightly underestimate the true EP
rate, which will not affect the validity of our trade-off relation.
It is clear that the information of microscopic states is still needed
to calculate the EPR using this expression, even
if the model details are known in prior.}

\section{\normalcolor{Simulation details and further numerical results}}

\subsection{\textit{\normalcolor{Calculating Lower Bounds for Interacting Brownian Particles with Different Levels of Coarse-Graining}}}

\begin{figure}
\begin{centering}
\normalcolor{\includegraphics[width=1\columnwidth]{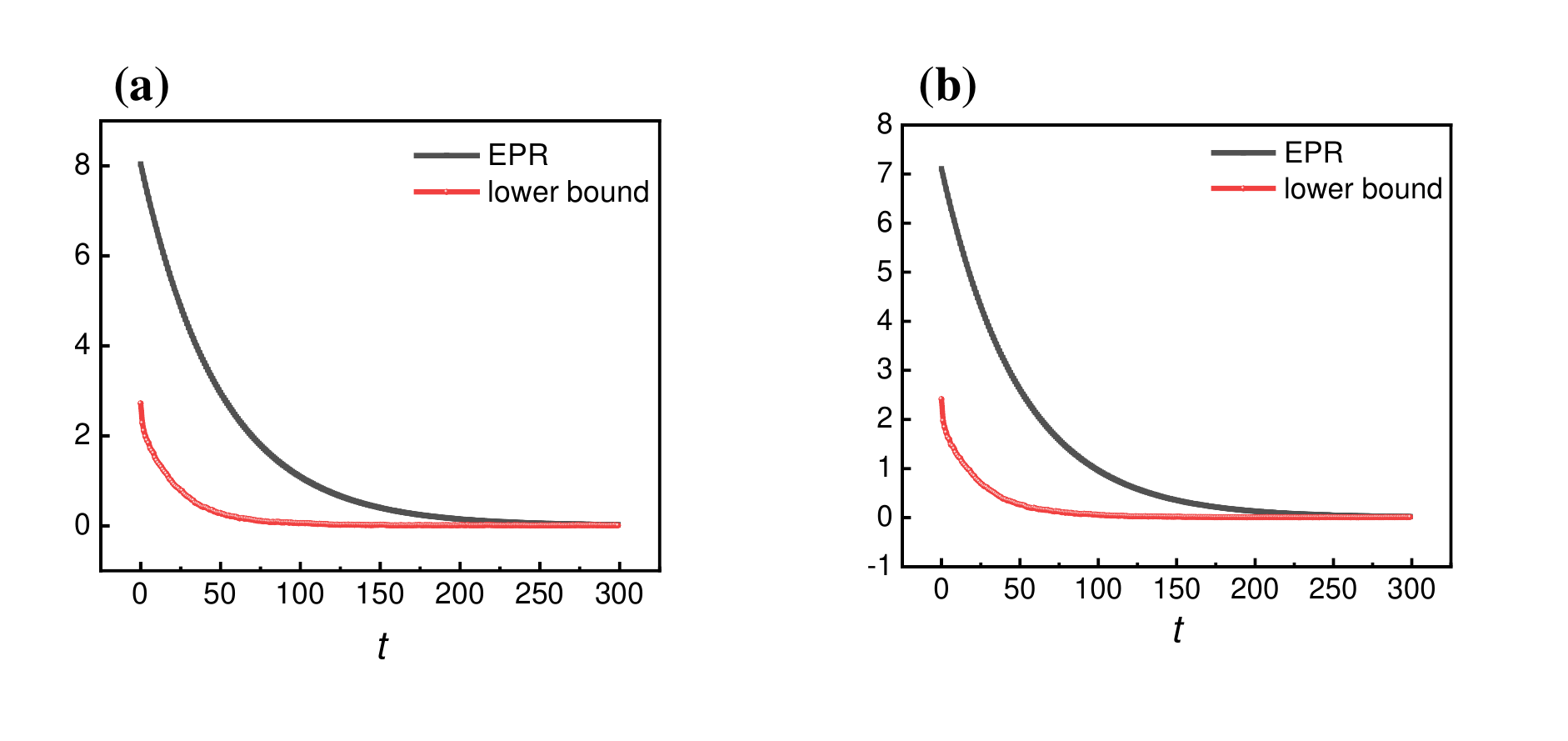}}
\par\end{centering}
\normalcolor{\caption{\normalcolor{The coarse-grained trade-off relations for the interacting
Brownian particles system with more coarse-grained states: (a) The
lower bound is calculated by dividing the space uniformly into 16
regions (16 states). (b) The lower bound is calculated by dividing
the space uniformly into 25 regions (25 states).}}
}

\normalcolor{\label{S1}}
\end{figure}

\normalcolor{The interaction potential is the spring potential,
which reads 
\begin{equation}
U_{in}(r_{ij})=\begin{cases}
\frac{1}{2}\kappa(r_{ij}-r_{c})^{2} & r_{ij}<r_{c}\\
0 & r_{ij}\geq r_{c}
\end{cases}.
\end{equation}
Here, the $r_{c}=1.0$ is the cutoff distance. $\kappa=0.01\ll k=0.1$,
where $k$ is the stiffness of the external field $U(\boldsymbol{x})=k||\boldsymbol{x}-\boldsymbol{a}||^{2}/2$.
The diffusion constant $D=k_{B}T=1.0$. The simulation box has a size
of $50\times50$ with periodic boundary conditions. In Fig. \ref{S1},
we divide the 2-dimensional box uniformly into $16$ and $25$ regions,
in contrast with the $4$ regions in the main text. When the coarse-grained
level increases (the number of coarse-grained states increases), the
lower bound becomes closer to the real EPR.}

\normalcolor{}

\subsection{\textit{\normalcolor{Another example: interacting active Brownian particles
system}}\normalcolor{{} }}

\normalcolor{We consider a more complex scenario involving interacting
active Brownian particles. The dynamics of this system is described
by the following overdamped Langevin equations
\begin{align}
\dot{\boldsymbol{r}} & = D\beta[\boldsymbol{F}_{i}+\boldsymbol{n}_{i}v]+\sqrt{2D}\boldsymbol{\xi}_{i}(t)\nonumber \\
\dot{\theta}_{i} & =\sqrt{2D_{r}}\eta_{i}(t),
\end{align}
where $\boldsymbol{F}_{i}=-\sum_{i\neq j}\nabla U(r_{ij})$ and $v$
denotes the strength of active force whose orientation is described
by the unit vector $\boldsymbol{n}_{i}=(\cos\theta_{i},\sin\theta_{i})$.
$\beta=\frac{1}{k_{B}T}=10.0$. The strength of active force is chosen to be $v=20.0$. More simulation details can be found in the reference \cite{du2019self}. We define $\boldsymbol{F}_{tot}\equiv \boldsymbol{F}_{i}+\boldsymbol{n}_{i}v$. With detailed knowledged of the model, the (non-adiabatic) EPR at time $t$ is calculated approximately by numerically integrating $\dot{\sigma}(t)=\frac{1}{\delta t}\int_{t}^{t+\delta t} \boldsymbol{F}_{tot}\circ \dot{\boldsymbol{r}} dt $, where $\delta t$ is a small time step.}

\normalcolor{}
\begin{figure}
\begin{centering}
\normalcolor{\includegraphics[width=1\columnwidth]{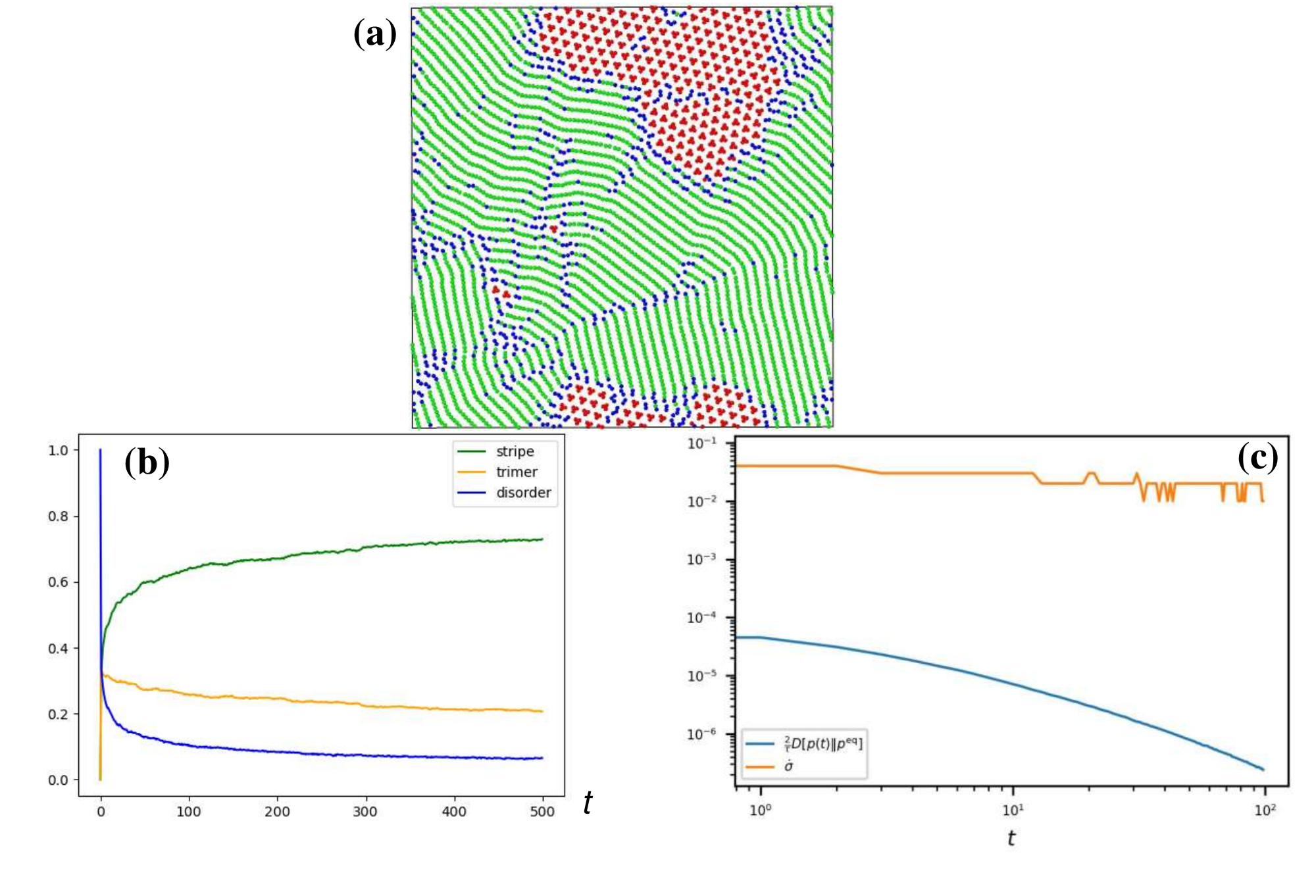}}
\par\end{centering}
\normalcolor{\caption{\normalcolor{Demonstration of the coarse-grained trade-off relation
in an interacting active particles system. There are 4096 active particles.
(a) A snapshot of the simulation, showing the system self-assembling
into three coarse-grained states: the stripe state (green part), the
trimer state (red part), and the disorder state (blue part). (b) The
evolution of the probability distributions of the three coarse-grained
states during relaxation. (c) The non-adiabatic EPR calculated from
detailed knowledge (orange curve) and the coarse-grained lower bound
(blue curve) using only the statistics of the coarse-grained states. }}
}

\normalcolor{\label{S2}}
\end{figure}

\normalcolor{This example demonstrates the limitation of our inference
strategy, specifically that it can only provide a very loose bound
when the accessible dynamics are highly coarse-grained. This remains
an open question within the stochastic thermodynamics community, as
the reduction of EP due to coarse-graining (information
loss) is inevitable. If we apply our lower bound to systems with more
coarse-grained states, we anticipate that the bound will more closely
approximate the true EPR, similar to the first
example.}

\section{Generalization to open quantum systems and continuous-space Markov
processes}

\subsection{\textit{Details of Generalization to Markovian open quantum systems} }

Here we show that, our main results can be generalized to quantum
Markov processes described by the Lindblad master equations. In this
setting, the dynamics of the density operator $\rho_{t}\equiv\rho(t)$
\normalcolor{of the system} at time $t$ is given by $\dot{\rho}_{t}=\mathcal{L}(\rho_{t}),$
where
\begin{equation}
\mathcal{L}_{t}(\rho)\equiv-i[H_{t},\rho]+\sum_{i}\left[J_{i}\rho J_{i}^{\dagger}-\frac{1}{2}\{J_{i}^{\dagger}J_{i},\rho\}\right]\label{Lindbladian}
\end{equation}
is the Lindbladian. Here, $H_{t}$ is the Hamiltonian \normalcolor{(can
possibly be time-dependent)} in a $d-$dimensional Hilbert space $\mathcal{H}^{d}$
and $J_{i}$ is the $j$th jump operator describing dissipation effect
due to the environment. To proceed, we assume that the Lindbladian
satisfies the quantum detailed balance condition \cite{firanko2022area},
in which case the density operator will finally converge to a Gibbs
state $\rho_{\beta}=\lim_{t\rightarrow\infty}e^{\mathcal{L}t}\rho_{0}=e^{-\beta H}/\text{Tr(\ensuremath{e^{-\beta H}})}$
\normalcolor{when $H$ is time-independent.} The inner product
should be redefined as $\langle A,B\rangle_{\pi}\equiv\text{Tr}(A^{\dagger}B\rho_{\beta})$,
the average over the Gibbs state reads $\langle A\rangle_{\pi}\equiv\text{Tr}(A\rho_{\beta})$
and the quantum KL divergence is given by $D(\rho||\rho^{\prime})=\text{Tr}(\rho\ln\rho-\rho\ln\rho^{\prime})$.
The EPR $\dot{\sigma}_{t}$ at time $t$ in the
open quantum systems can be separated to the change rate of system
entropy $\dot{s}_{t}=\text{Tr}(\dot{\rho_{t}}\ln\rho_{t})$ and the
heat flow $\beta\dot{q}=\beta\text{Tr(\ensuremath{\dot{\rho}_{t}H_{t}})}$
as $\dot{\sigma}_{t}=\dot{s}_{t}-\beta\dot{q}$. Like in the classical
case, $\dot{\sigma}_{t}$ has a direct connection with KL divergence
that $\dot{\sigma}_{t}=-\partial_{t}D(\rho_{t}||\rho_{\beta})$ \cite{breuer2002theory}.
\normalcolor{When $H$ is time-dependent, $\dot{\sigma}_{t}=-\partial_{t}D(\rho_{t}||\rho_{\beta,\tau})\vert_{t=\tau}$,
where $\rho_{\beta,\tau}$ is the instantaneous Gibbs state defined
as $\mathcal{L}_{\tau}(\rho_{\beta,\tau})=0$. }Recently, it has been
proved that there always exist a positive constant $\alpha$ assuring
that the quantum LS inequality 
\begin{equation}
-\langle\mathcal{L}(f),\ln f\rangle_{\pi}\geq\alpha\langle f\ln f\rangle_{\pi}
\end{equation}
holds \cite{gao2022complete} for any postive operator $f\in\mathcal{H}^{d}$
satisfying $\langle f\rangle_{\pi}=1$, once the quantum detailed
balance condition is satisfied. Consequently, a straightforward calculation
shows that, 
\begin{align}
\dot{\sigma}_{t}= & -\partial_{t}D(\rho_{t}||\rho_{\beta,\tau})\vert_{t=\tau}\nonumber \\
= & -\langle\mathcal{L}(\frac{\rho_{t}}{\rho_{\beta,t}}),\ln\frac{\rho_{t}}{\rho_{\beta,t}}\rangle_{\pi}(t)\nonumber \\
\geq & \lambda_{\text{QLS}}(t)\langle\frac{\rho_{t}}{\rho_{\beta,t}}\ln\frac{\rho_{t}}{\rho_{\beta,t}}\rangle_{\pi},\nonumber \\
= & \lambda_{\text{QLS}}(t)D(\rho_{t}||\rho_{\beta,t}),\label{QLS}
\end{align}
where 
\begin{equation}
\lambda_{\text{QLS}}\equiv\inf_{f>0,\ \langle f\rangle_{\pi}=1}\frac{-\langle\mathcal{L}(f),\ln f\rangle_{\pi}}{\text{\ensuremath{\langle f\ln f\rangle_{\pi}}}}>0
\end{equation}
is the quantum LS constant. Integrating both parts of Eq. (\ref{QLS})
from $0$ to $t$ results in $D(\rho_{t}||\rho_{\beta})\leq D(\rho_{0}||\rho_{\beta})e^{-\lambda_{\text{QLS}}t}$.
Then, an inverse quantum speed limit can still be directly obtained
as in the classical case when $H$ is time-independent: 
\begin{equation}
\tau\leq\frac{1}{\lambda_{\text{QLS}}}\ln\left\{ \frac{\sigma_{\text{tot}}^{0}}{\sigma_{\text{tot}}^{0}-\sigma_{[0,\tau]}}\right\} ,
\end{equation}
where $\sigma_{[0,\tau]}=\int_{0}^{\tau}\dot{\sigma}_{t}\text{d}t=D(\rho_{0}||\rho_{\beta})-D(\rho_{t}||\rho_{\beta})$
and $\sigma_{\text{tot}}^{0}=D(\rho_{0}||\rho_{\beta})$. The quantum
LS constant can also be connected with the relaxation time scale by
using the Pinsker inequality $(\text{Tr}\vert\rho-\rho^{\prime}\vert)^{2}\leq2D(\rho||\rho^{\prime})$.
In open quantum systems, the distance between two density operator
is commonly described by the trace distance defined as $D_{\text{Tr}}(\rho||\rho^{\prime})\equiv\frac{1}{2}\text{Tr}\vert\rho-\rho^{\prime}\vert$.
Thus the relaxation time scale to the Gibbs state is naturally characterized
by the convergence rate of $D_{\text{Tr}}(\rho_{t}||\rho_{\beta})$.
Here, we have that
\begin{align}
D_{\text{Tr}}(\rho_{t}||\rho_{\beta})\leq & \sqrt{D(\rho_{t}||\rho_{\beta})/2}\nonumber \\
\leq & \sqrt{D(\rho_{0}||\rho_{\beta})/2}e^{-\lambda_{\text{QLS}}t/2},
\end{align}
which implies that $\lambda_{\text{QLS}}$ is a characterization of
relaxation time scale in open quantum systems. In summary, the upper
bound of the transformation time $\tau$ in relaxation of open quantum
systems depends both on the relaxation time scale of the whole process
and the initial energetic cost, similar to the classical case.

\normalcolor{In addition, we provide an alternative way to generalize
our results to open quantum systems when $H$ is time-independent,
which is the generalization shown in the Eq. (10) of main text. As
is known, the Lindblad master equation can lead to an equation of
motion for the populations 
\begin{equation}
P_{n}(t)\equiv\langle n\vert\rho_{t}\vert n\rangle
\end{equation}
of the eigenstates $\vert n\rangle$ of the system Hamiltonian $H_{s}=\sum_{n}\epsilon_{n}\vert n\rangle\langle n\vert$
(assuming that $H_{s}$ is non-degenerate) \cite{breuer2002theory}.
The equation of motion is given by 
\begin{equation}
\frac{\text{d}P_{n}(t)}{\text{d}t}=\sum_{m}\left[W_{nm}P_{m}(t)-W_{mn}P_{n}(t)\right],
\end{equation}
which is often referred to as the Pauli master equation. In this equation,
time-independent transition rates $W_{nm}$ from the energy level
$\epsilon_{m}$ to $\epsilon_{n}$ are given by 
\[
W_{nm}=\sum_{i}(\epsilon_{m}-\epsilon_{n})\langle n\vert J_{i}\vert m\rangle^{2}.
\]
If the LS constant associated with the Markov generator $\tilde{\boldsymbol{W}}$
(whose entries are $\tilde{W}_{nm}=W_{nm}-\delta_{nm}\sum_{l}W_{ln}$)
is $\lambda_{\text{LS}}$, one has that 
\begin{align}
-\frac{\text{d}}{\text{d}t}D[\boldsymbol{P}(t)\vert\vert\boldsymbol{P_{\beta}}] & \geq4\lambda_{\text{LS}}D[\boldsymbol{P}(t)\vert\vert\boldsymbol{P_{\beta}}]\\
D[\boldsymbol{P}(\tau)\vert\vert\boldsymbol{P_{\beta}}] & \leq e^{-4\lambda_{\text{LS}}\tau}D[\boldsymbol{P}(\tau)\vert\vert\boldsymbol{P_{\beta}}],
\end{align}
where $D[\cdot\vert\vert\cdot]$ is the classical KL divergence and
entries of $\boldsymbol{P_{\beta}}$ are given by $P_{\beta,n}=\langle n\vert\rho_{\beta}\vert n\rangle$.
The quantum EP can be decomposed as 
\begin{equation}
\dot{\sigma}_{t}=-\partial_{t}D(\rho_{t}||\rho_{\beta})=-\frac{\text{d}}{\text{d}t}D[\boldsymbol{P}(t)\vert\vert\boldsymbol{P_{\beta}}]-\frac{\text{d}A(t)}{\text{d}t},
\end{equation}
where $A(t)$ is the asymmetry defined as $A(t):=D(\rho_{t}||\rho_{t}^{\text{d}})$.
Here, $\rho_{t}^{\text{d}}$ is the fully decohered version of $\rho_{t}$.
It has been shown that $-\frac{\text{d}A(t)}{\text{d}t}\geq0$, which
quantifies the EPR from destroying quantum coherence.
Consequently, we have that 
\begin{equation}
\dot{\sigma}_{t}\geq4\lambda_{\text{LS}}D[\boldsymbol{P}(t)\vert\vert\boldsymbol{P_{\beta}}]-\frac{\text{d}A(t)}{\text{d}t}\geq4\lambda_{\text{LS}}D[\boldsymbol{P}(t)\vert\vert\boldsymbol{P_{\beta}}],
\end{equation}
and the EP 
\begin{equation}
\sigma_{[0,\tau]}=D[\boldsymbol{P}(0)\vert\vert\boldsymbol{P_{\beta}}]-D[\boldsymbol{P}(\tau)\vert\vert\boldsymbol{P_{\beta}}]+A(0)-A(\tau)
\end{equation}
during any interval $[0,\tau]$ is also bounded from below as
\begin{equation}
\sigma_{[0,\tau]}\geq(1-e^{-4\lambda_{\text{LS}}\tau})D[\boldsymbol{P}(0)\vert\vert\boldsymbol{P_{\beta}}]+\Delta A,
\end{equation}
with $\Delta A\equiv A(0)-A(\tau)\geq0$. Thus, a quantum inverse
speed limit tighter than its classical counterpart is obtained as
\begin{equation}
\tau\leq\frac{1}{4\lambda_{\text{LS}}}\ln\left\{ \frac{D[\boldsymbol{P}(0)\vert\vert\boldsymbol{P_{\beta}}]}{D[\boldsymbol{P}(0)\vert\vert\boldsymbol{P_{\beta}}]-(\sigma_{[0,\tau]}-\Delta A)}\right\} .
\end{equation}
}

Whether the above relations hold true in the nonequilibrium open quantum
systems remains an interesting open problem.

\subsection{\textit{Generalization to Continuous-space Markov processes}}

The Markov processes in continuous-space can be described by the Fokker-Planck
equation. Here, we would like to discuss a system within a time-dependent
conservative force field $U(x,t)$, where $x\in\mathbb{R}^{n}$ is
a $n$-dimensional vector. The dynamics of the system is described
by a Langevin equation (we have set the mobility $\mu=1$)
\begin{equation}
\dot{x}(t)=F(x,t)+\sqrt{2D}\xi(t),
\end{equation}
where the force $F(x,t)\equiv-\nabla U(x,t)$. The corresponding Fokker-Planck
equation reads
\begin{equation}
\frac{\partial p(x,t)}{\partial t}=-\nabla j(x,t),\label{FPE}
\end{equation}
where the current $j(x,t)\equiv F(x,t)p(x,t)-D\nabla p(x,t)$. The
Fokker-Planck equation has an instantaneous stationary solution $p_{t}^{st}(x)\propto e^{-\beta U(x,t)}$
with Boltzmann form at any time $t$. 

According to Ref. \cite{2000trend_Villani}, the LS inequality
\begin{equation}
\frac{\lambda}{2}\int u\ln up^{st}(x)\text{d}x\leq\int\vert\nabla\sqrt{u}\vert^{2}p^{st}(x)\text{d}x\label{LSFPE}
\end{equation}
holds for any positive function $u=u(x,t)$ satisfying $\int u(x,t)p^{st}(x)\text{d}x=1$
and any stationary distribution $p^{st}(x)\propto e^{-\beta U(x)}$,
if the following condition is fulfilled for the positive constant
$\lambda$:
\[
\nabla^{2}U(x)\geq\lambda\mathbb{I}_{n},
\]
where $\mathbb{I}_{n}$ is the $n$-dimensional identity matrix. For
instance, consider a one-dimensional Brownian particle confined in
a harmonic potential $U(x)=\frac{1}{2}kx^{2}$. In this case, $U^{\prime\prime}(x)=k$,
such that the positive constant $\lambda$ is exactly equal to the
stiffness $k$ of the potential. 

The EPR $\dot{\sigma}(t)$ at time $t$ obtained
from Eq. (\ref{FPE}) can be expressed as the time-derivative of the
KL divergence \cite{22PRE_ItoCoupling}, i.e., 
\begin{equation}
\dot{\sigma}(t)=-\frac{\text{d}}{\text{d}t}D[p(x,s)||p_{t}^{st}(x)]\vert_{s=t}=\int\left|\nabla\ln\left[\frac{p(x,t)}{p_{t}^{st}(x)}\right]\right|^{2}p(x,t)\text{d}x\geq0.
\end{equation}
Notice that $\dot{\sigma}(t)$ can be rewritten as 
\begin{equation}
\dot{\sigma}(t)=4\int\left|\nabla\sqrt{\frac{p(x,t)}{p_{t}^{st}(x)}}\right|^{2}p_{t}^{st}(x)\text{d}x,
\end{equation}
then applying the LS inequality (\ref{LSFPE}) to it yields 
\begin{equation}
\dot{\sigma}(t)\geq2\lambda\int\frac{p(x,t)}{p_{t}^{st}(x)}\ln\frac{p(x,t)}{p_{t}^{st}(x)}p^{st}(x)\text{d}x=2\lambda D[p(x,t)||p_{t}^{st}(x)].
\end{equation}
Therefore, there is still a general lower bound $\dot{\sigma}(t)\geq2\lambda D[p(x,t)||p_{t}^{st}(x)]$
for the continuous-space Markov process described by Eq. (\ref{FPE}),
serving as a stronger second law of thermodynamics.

Whether it is possible to find a general lower bound for the local
EPR using local version of the KL divergence or
other distance function is another interesting open question. 

\subsection{\textit{Generalization to discrete-time Markov processes}}

The non-adiabatic EP for such process during relaxation
can also be expressed by KL divergence as
\begin{equation}
\sigma_{[0,\tau]}=D[\boldsymbol{p}_{0}\vert\vert\boldsymbol{\pi}]-D[\boldsymbol{p}_{\tau}\vert\vert\boldsymbol{\pi}].
\end{equation}
It can be proved that 
\begin{equation}
D[\boldsymbol{p}_{\tau}\vert\vert\boldsymbol{\pi}]\leq\left(1-\alpha_{d}\right)^{\tau}D[\boldsymbol{p}_{0}\vert\vert\boldsymbol{\pi}],
\end{equation}
leading to the result 
\begin{equation}
\sigma_{[0,\tau]}\geq\left[1-\left(1-\alpha_{d}\right)^{\tau}\right]D[\boldsymbol{p}_{0}\vert\vert\boldsymbol{\pi}],
\end{equation}
where the LS constant $\alpha_{d}$ for discrete-time case is defined
as 
\[
\alpha_{d}\equiv\min\left\{ \frac{\langle(1-KK^{\star})f,f\rangle}{\text{Ent}(f)}\right\} .
\]

\end{document}